\documentclass[twocolumn,prd,amssymb,aps,nofootinbib,showpacs,epsf]{revtex4} 
\usepackage[usenames]{color} 
\usepackage{ulem} 
\usepackage{graphicx}
\usepackage{amssymb}
\usepackage{amsmath}
\usepackage{epstopdf}
\usepackage{aas_macros}
\newcommand{\bbe}{\begin{equation}}
\newcommand{\be}{\begin{equation}}
\newcommand{\ee}{\end{equation}}
\newcommand{\beaa}{\begin{eqnarray*}}
\newcommand{\eeaa}{\end{eqnarray*}}
\newcommand{\ben}{\begin{enumerate}}
\newcommand{\een}{\end{enumerate}}
\newcommand{\dis}{\displaystyle}
\begin{document}
\title{Constraints on a phenomenologically parameterized neutron-star
equation of state}
\author{Jocelyn S. Read$^1$, Benjamin D. Lackey$^1$, Benjamin J. Owen$^2$,
and John L. Friedman$^1$}
\affiliation{
$^1$Department of Physics, University of Wisconsin-Milwaukee, P.O. Box
413, Milwaukee, WI 53201
\\
$^2$Center for Gravitational Wave Physics, Institute for Gravitation and
the Cosmos, and Department of Physics, The Pennsylvania State University,
University Park, PA 16802-6300
}

\begin{abstract}
We introduce a parameterized high-density equation of state (EOS) in order to
systematize the study of constraints placed by astrophysical observations on
the nature of neutron-star matter.  To obtain useful constraints, the number of
parameters should be smaller than the number of neutron-star properties that
have been measured or will have been measured in the next several years.  And
the set must be large enough to accurately approximate the large set of
candidate EOSs.  We find that a parameterized EOS based on piecewise polytropes
with 3 free parameters matches to about 4\% rms error an extensive set of
candidate EOSs at densities below the central density of $1.4\, M_\odot$ stars.
Adding observations of more massive stars constrains the higher density part of
the EOS and requires an additional parameter.  
We obtain constraints on the
allowed parameter space set by causality and by present and near-future
astronomical observations.  
In particular, we emphasize potentially stringent 
constraints on the EOS parameter space associated with two measured properties 
of a single star;  and we find that a measurement of the moment of
inertia of PSR~J0737-3039A can strongly constrain the maximum
neutron-star mass.
We also present in an appendix a more efficient algorithm 
than has previously been used for finding points of marginal stability 
and the maximum angular velocity of stable stars.
\end{abstract}


\maketitle

\section{Introduction}

Because the temperature of neutron stars is far below the Fermi energy of their
constituent particles, neutron-star matter is accurately described by the
one-parameter equation of state (EOS) that governs cold matter above nuclear
density.  The
uncertainty in that EOS, however, is notoriously large, with the pressure $p$
as a function of baryon mass density $\rho$ uncertain by as much as an order of
magnitude above nuclear density.  The phase of the matter in the core of a
neutron star is similarly uncertain: Current candidates for the EOS include
non-relativistic and relativistic mean-field models; models for which
neutron-star cores are dominated by nucleons, by hyperons, by pion or kaon
condensates, and by strange quark matter (free up, down, and strange quarks);
and one cannot yet rule out the possibility that the ground state
of cold matter at zero pressure might be strange quark matter and that the term
``neutron star'' is a misnomer for strange quark stars.  

The correspondingly large number of fundamental parameters needed to
accommodate the models' Lagrangians has meant that studies of astrophysical
constraints (see, for example, \cite{EngvikOsnesHjorthJensen1996constr,
Lattimer:2000nx, lattimerprakash2006nuc,
KlahnBlaschkeTypel2006constraints, PageReddy2006review} and references therein)
present constraints by dividing the EOS candidates into an allowed and a
ruled-out list.  A more systematic study, in which astrophysical constraints
are described as constraints on the parameter space of a parameterized EOS, can
be effective only if the number of parameters is smaller than the number of
neutron-star properties that have been measured or will have been measured in
the next several years.  At the same time, the number of parameters must be
large enough to accurately approximate the EOS candidates.  

A principal aim of this paper is to show that, if one uses phenomenological
rather than fundamental parameters, one can obtain a parameterized EOS that
meets these conditions.  We exhibit a parameterized EOS, based on specifying
the stiffness of the star in three density intervals, measured by the adiabatic
index $\Gamma = d\log P/d\log\rho$.  A fourth parameter translates the
$p(\rho)$ curve up or down, adding a constant pressure -- equivalently fixing
the pressure at the endpoint of the first density interval.  Finally, the EOS
is matched below nuclear density to the (presumed known) EOS. An EOS for which
$\Gamma$ is constant is a polytrope, and the parameterized EOS is then
piecewise polytropic.  A similar piecewise-polytropic EOS was previously
considered by Vuille and Ipser \cite{VuilleIpser1999max}; and, with different
motivation, several other authors \cite{ZdunikBejgerHaensel2006phase,
BejgerHaenselZdunik2005phase, HaenselPotekhin2004anal, shibata2005es} have used
piecewise polytropes to approximate neutron-star EOS candidates.
In contrast to this previous work, we use a small 
number of parameters chosen to fit a wide variety of fundamental EOSs, 
and we systematically explore a variety of astrophysical constraints.
Like most of the previous work, we aim to model equations of state
containing nuclear matter (possibly with various phase transitions) rather
than pure quark stars, whose EOS is predicted to be substantially
different.

As we have noted, enough uncertainty remains in the pressure at nuclear
density, that one cannot simply match to a fiducial pressure at
$\rho_{\mathrm{nuc}}$.  Instead of taking as one parameter the pressure at a
fiducial density, however, one could match to the pressure of the known
subnuclear EOS at, say, 0.1 $\rho_{\mathrm{nuc}}$ and then use as one parameter
a value of $\Gamma_0$ for the interval between $0.1\rho_{\mathrm{nuc}}$ and
$\rho_{\mathrm{nuc}}$.  Neutron-star observables are insensitive to the EOS
below $\rho_{\mathrm{nuc}}$, because the fraction of mass at low density is
small.  But the new parameter $\Gamma_0$ would indirectly affect observables by
changing the value of the pressure at and above nuclear density, for fixed
values of the remaining $\Gamma_i$.  By choosing instead the pressure at a
fixed density $\rho_1 > \rho_{\mathrm{nuc}}$, we obtain a parameter more
directly connected to physical observables.  In particular, as
Lattimer and Prakash~\cite{Lattimer:2000nx} have pointed out, neutron-star
radii are closely tied to the pressure somewhat above nuclear density, and
the choice $p_1 = p(\rho_1)$ is recommended by that relation. 

In general, to specify a piecewise polytropic EOS with three density intervals
above nuclear density, one needs six parameters: two dividing densities, three
adiabatic indices $\Gamma_i$, and a value of the pressure at an endpoint of one
of the intervals.  Remarkably, however, we find (in Sec.~\ref{sec:fits}) that
the error in fitting the collection of EOS candidates has a clear minimum for a
particular choice of dividing densities.  With that choice, the parametrized
EOS has three free parameters, $\Gamma_1, \Gamma_2$ and $p_1$, for densities
below $10^{15}$~g/cm$^3$ (the density range most relevant for masses $\sim 1.4
M_\odot$), and four free parameters (an additional $\Gamma_3$) for
densities between $10^{15}$~g/cm$^3$ and the central density of the maximum
mass star for each EOS.  

With the parameterization in hand, we examine in Sect.~\ref{sec:constraints}
astrophysical constraints on the parameter space beyond
the radius-$p_1$ relation found by Lattimer and
Prakash~\cite{Lattimer:2000nx}.  Our emphasis in this first study is on
present and very near-future constraints:  those associated with the
largest observed neutron-star mass and spin, with a possible observation (as
yet unrepeated) of neutron-star redshift,
with a possible simultaneous measurement of mass and
radius,
and with the expected future
measurement of the moment of inertia of a neutron star with known mass.
A companion paper~\cite{nrda} will investigate constraints
obtainable with gravitational-wave observations in a few years.
The
constraints associated with  the largest observed mass, spin, and redshift have
a similar form, each restricting  the parameter space to one side of a surface:
For example, if we take the largest  observed mass (at a 90\% confidence level)
to be 1.7 $M_\odot$, then the allowed  parameters correspond to EOSs whose
maximum mass is at least 1.7 $M_\odot$.  We can regard $M_{\rm max}$ as a
function on the 4-dimensional EOS parameter space. The subspace of EOSs for
which $M_{\rm max} = 1.7 M_\odot$ is then described by a  3-dimensional
surface, and constraint is a restriction to the high-mass side of the surface.
Similarly, the observation of a 716 Hz pulsar restricts the EOS parameter space
to one side of a surface that describes EOSs for which the maximum spin is  716
Hz. 
Thus we can produce model-independent extended versions of
the multidimensional constraints seen in~\cite{Lackey:2005tk}.

The potential simultaneous observation of two properties of a single neutron
star (for example, moment of inertia and mass) would yield a significantly
stronger constraint: It would restrict the parameter space not to one side of a
surface but to the surface itself. And a subsequent observation of two
different parameters for a different neutron star would then restrict one to
the intersection of two surfaces. We exhibit the  result of simultaneous
observations of mass and moment of inertia (expected within  the next decade
for one member of the binary pulsar J0737-3039~\cite{Lattimer:2004nj,
bejger-bulik-haensel}) and  of mass and radius.  Gravitational-wave
observations of binary inspiral can again  measure two properties of a single
star: both mass and an observable roughly described as the final frequency
before plunge (the departure of the waveform from a point-particle inspiral);
and related work in progress examines the accuracy with which one can extract
EOS parameters from interferometric observations of gravitational waves in the
inspiral and merger of a binary neutron star system \cite{nrda}.  

{\sl Conventions}:  We use cgs units, denoting rest-mass density by $\rho$, and
(energy density)/$c^2$ by $\epsilon$.  We define rest-mass density as
$\rho=m_{\rm B}n$ where $m_{\rm B}=1.66\times 10^{-24}$~g and $n$ is the
baryon number density. In Sec.~\ref{sec:ppdef}, however, we set $c=1$ to
simplify the equations and add a footnote on restoring $c$.

\section{Candidates}
\label{sec:candidates}

A test of how well a parametrized EOS can approximate the true EOS of cold
matter at high density is how well it approximates candidate EOSs.
We consider a wide array of candidate EOSs, covering many different generation methods
and potential species. 
Because the parametrized EOS is intended to distinguish the parts of
parameter space allowed and ruled out by present and future observations,
the collection includes some EOSs that no longer satisfy known
observational constraints.  Many of the candidate EOSs were considered in
Refs.~\cite{Lattimer:2000nx, bejger-bulik-haensel, Lackey:2005tk}; and we
call them by the names used in those papers.

For plain $npe\mu$  nuclear matter, we include:
\begin{itemize}
\item Two potential-method EOSs (PAL6~\cite{pal} and SLy~\cite{sly4}), 
\item eight variational-method EOSs (AP1-4~\cite{apr}, FPS~\cite{fp}, and WFF1-3~\cite{wff}), 
\item one nonrelativistic (BBB2~\cite{bbb2}) and three relativistic
  (BPAL12~\cite{bpal12}, ENG~\cite{engvik} and MPA1~\cite{mpa1})
Brueckner-Hartree-Fock EOSs, and
\item three relativistic mean field theory EOSs (MS1-2 and one we call MS1b,
which is identical to MS1 except with a low symmetry energy of 25~MeV~\cite{ms}).
\end{itemize}

We also consider models with hyperons, pion and kaon condensates, and quarks,
and will collectively refer to these EOSs as $K/\pi/H/q$ models.
\begin{itemize}
\item one neutron-only EOS with pion condensates (PS~\cite{ps}),
\item two relativistic mean field theory EOSs with kaons
  (GS1-2~\cite{schaf}), 
\item one effective potential EOS including hyperons (BGN1H1~\cite{balbn1h1}),
\item eight relativistic mean field theory EOSs with hyperons
(GNH3~\cite{glendnh3} and seven variants H1-7~\cite{Lackey:2005tk},
\item one relativistic mean field theory EOS with hyperons and quarks
(PCL2~\cite{pcl}), and
\item four hybrid EOSs with mixed APR nuclear matter and colour-flavor-locked
quark matter (ALF1-4 with transition density $\rho_{\rm c}$ and
interaction parameter $c$ given by $\rho_{\rm c}=2n_0$,  $c=0$;
$\rho_{\rm c}=3n_0$, $c=0.3$; $\rho_{\rm c}=3n_0$, $c=0.3$; and
$\rho_{\rm c}=4.5n_0$, 
$c=0.3$ respectively~\cite{AlfordBraby2005hybrid}).
\end{itemize}

\begin{figure}[!htb]
\includegraphics[width=3.4in]{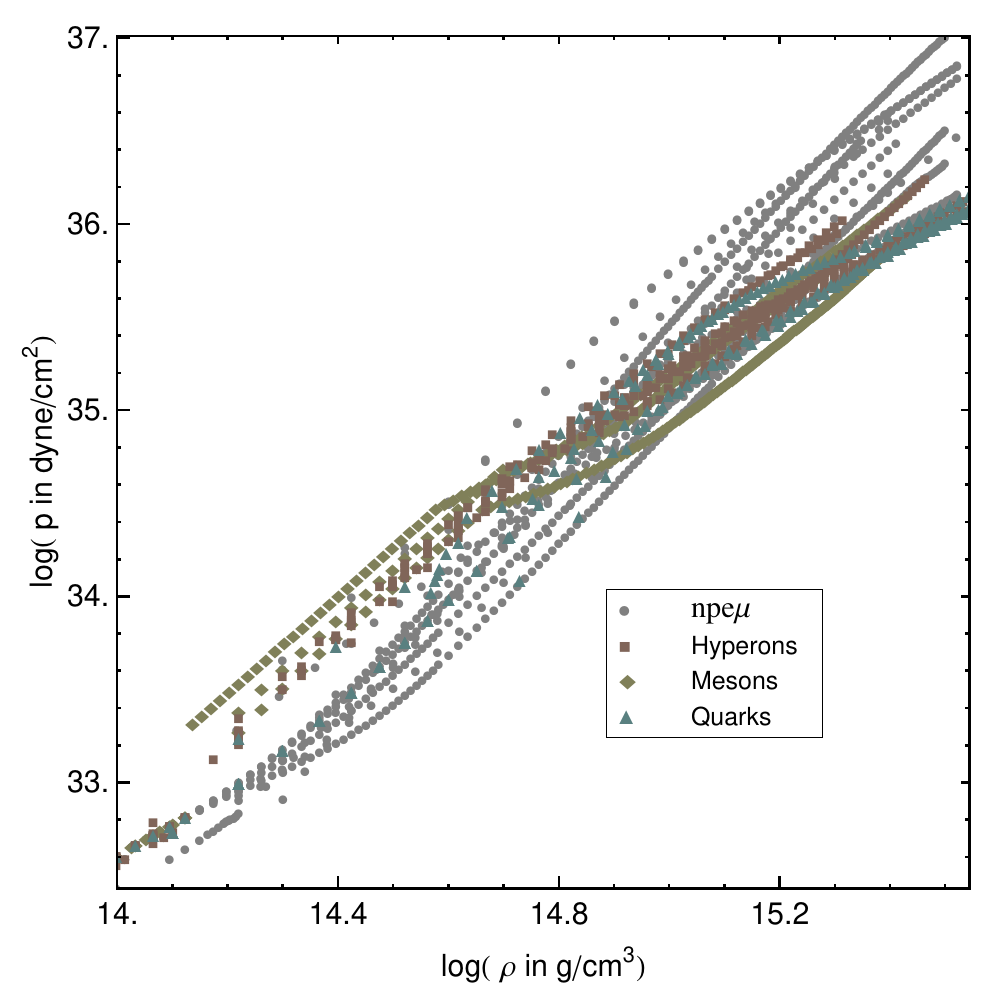}
\caption{
Pressure versus rest mass density for the set of candidate EOS
tables considered in the parameterization.}
\label{fig:candrange}
\end{figure}

The tables are plotted in Fig.~\ref{fig:candrange} to give an idea of the range
of EOSs considered for this parameterization.

\section{Piecewise polytrope}
\label{sec:ppdef}

A polytropic EOS has the form, 
\begin{equation}
p(\rho) = K \rho^{\Gamma},
\label{eq:polyp}\end{equation}
with $\rho$ the rest-mass density and $\Gamma$ the adiabatic index,  and with
energy density 
$\epsilon$ fixed by the first law of thermodynamics,\footnote{In this section, 
for simplicity of notation, $c=1$.  To rewrite the equations in cgs units, 
replace $p$ and $K$ in each occurrence by $p/c^2$ and $K/c^2$. Both 
$\epsilon$ and $\rho$ have units g/cm$^3$.\hfill}
\be
 d\frac\epsilon\rho= -p\ d\frac1\rho.
\label{eq:1stlaw}\ee  
For $p$ of the form 
(\ref{eq:polyp}), Eq.~(\ref{eq:1stlaw}) has the immediate integral 
\be
  \frac\epsilon\rho= (1+a)+ \frac1{\Gamma -1}K\rho^{\Gamma-1}, 
\label{eq:polyep1}\ee
where $a$ is a constant; and the requirement 
$\dis \lim_{\rho \rightarrow 0}\frac{\epsilon}{\rho} = 1$ implies $a=0$  
and the standard relation 
\be
   \epsilon = \rho + \frac1{\Gamma -1} p.
\label{eq:polyep2}\ee
%A polytropic index $N$ is defined by $N = 1/(\Gamma-1)$.
 
The parameterized EOSs we consider are piecewise polytropes above a density
$\rho_0$, satisfying Eqs.~(\ref{eq:polyp}) and (\ref{eq:polyep1}) on a sequence
of density intervals, each with its own $K_i$ and $\Gamma_i$: An EOS is
piecewise polytropic for $\rho\geq\rho_0$ if, for a set of dividing densities
$\rho_0<\rho_1<\rho_2<\cdots $, the pressure and energy density are everywhere
continuous and satisfy 
\begin{equation}
p(\rho) = K_i \rho^{\Gamma_i}, \quad d\frac\epsilon\rho= -p d\frac1\rho,
 \quad \rho_{i-1} \leq \rho \leq \rho_i. 
\end{equation}
%\begin{equation}
%p(\rho) = \left\{ \begin{array}{ll}
%  K_0 \rho^{\Gamma_0} & \rho < \rho_1 \\
%  K_1 \rho^{\Gamma_1} & \rho_1 < \rho < \rho_2 \\
%    \vdots & \\
%  K_m \rho^{\Gamma_m} & \rho > \rho_m
%  \end{array} \right. 
%\end{equation}
%where $p$ is pressure, $\rho$ is rest mass density, and the $\Gamma_i$ are
%adiabatic indices.
Then, for $\Gamma \neq 1$,
\begin{equation}
\label{eq:endens}
\epsilon(\rho) = (1+ a_i) \rho + \frac{K_i}{\Gamma_i - 1} \rho^{\Gamma_i},
\end{equation}
with 
\begin{equation}
a_{i} = \frac{\epsilon(\rho_{i-1})}{\rho_{i-1}} - 1  - \frac{K_{i}}{\Gamma_{i} - 1}
\rho_{i-1}^{\Gamma_{i} - 1}.  
\end{equation}

The {\it specific enthalpy}%  
\footnote{A note on terminology:  
When the entropy vanishes, the specific 
enthalpy, $h=(\epsilon+p)/\rho$, and Gibbs free energy, 
$g=(\epsilon+p)/\rho-Ts$, coincide. For nonzero entropy, it is the term 
$g dM_0$ or, equivalently, $\mu dN$ that appears in the first law of 
thermodynamics, where $\mu=g/m_{\rm B}$ is the chemical potential.
Because $h=(\epsilon+p)/\rho$ 
always has the meaning of enthalpy, and because for isentropic stars 
and isentropic flows the constancy of injection energy and Bernoulli's law, 
respectively, are commonly stated in terms of enthalpy 
(see, for example \cite{pippard}), we refer to $(\epsilon+p)/\rho$ 
as the specific enthalpy, rather 
than the specific Gibbs free energy or the chemical potential.}
$h$ is defined as $(\epsilon + p)/\rho$ and is given 
in each density interval by
\begin{equation}
h(\rho) = 1 + a_i + \frac{\Gamma_i}{\Gamma_i - 1} K_i \rho^{\Gamma_i - 1}.
\end{equation}
The internal energy $e = \frac{\epsilon}{\rho} - 1$ is then
\begin{equation}
e(\rho) = a_i + \frac{K_i}{\Gamma_i - 1}\rho^{\Gamma_i - 1}
\end{equation}
and the sound velocity $v_{\rm s}$ is
\begin{equation}
v_{\rm s}(\rho) = \sqrt{\frac{dp}{d\epsilon}} 
	= \sqrt{\frac{\Gamma_i p}{\epsilon+p}}
\label{eq:vs}
\end{equation}

%When calculating in terms of enthalpy, it is useful to work with the
%quantity $\eta = H - 1$, as the expressions for $\rho$, $p$, and $\epsilon$
%in terms of $H$ involve $H-1$, and $H$ is near 1 for low densities.

Each piece of a piecewise polytropic EOS is specified by three parameters: the
initial density, the coefficient $K_i$, and the adiabatic index $\Gamma_i$.
However, when the EOS at lower density has already been specified up to the
chosen $\rho_i$, continuity of pressure restricts $K_{i+1}$ to the value
\begin{equation}
\label{eq:Kcont}
K_{i+1} = \frac{p(\rho_i)}{\rho_i^{\Gamma_{i+1}}}
\end{equation}
Thus each additional region requires only two additional parameters, $\rho_i$
and $\Gamma_{i+1}$.  Furthermore, if the initial density of an interval is
chosen to be a fixed value for the parameterization, specifying the EOS on the
density interval requires only a single additional parameter.

\section{Fitting methods}
\label{sec:fit-method}

As noted above, in choosing a parameterization for the EOS space, our goal was
to maintain high enough resolution, with a small number of parameters, that a
point in parameter space can accurately characterize the EOS of neutron-star
matter.  A measure of this resolution is the precision with which a point in
parameter space can fit the available candidate EOSs.  We describe in this
section how that measure is defined and computed for a choice of parameter
space -- that is, for a choice of the set of free parameters used to specify
piecewise polytropes.   
  
There is general agreement on the low-density EOS for cold matter, and we adopt
the version (SLy) given by Douchin and Haensel \cite{sly4}.  Substituting an
alternative low-density EOS from, for example, Negele and Vautherin
\cite{negele-vautherin}, alters by only a few percent the observables we
consider in examining astrophysical constraints, both because of the rough
agreement among the candidate EOSs and because the low density crust
contributes little to the mass, moment of inertia, or radius of the star. 

Each choice of a piecewise polytropic EOS above nuclear density is matched to
this low-density EOS.  The way in which the match is done is arbitrary, and,
again, the small contribution of the low-density crust to stellar observables
means that the choice of match does not significantly alter the relation
between astrophysical observables and the EOS above nuclear density. In our
choice of matching method, the first (lowest-density) piece of the piecewise
polytropic $p(\rho)$ curve is extended to lower densities until it intersects
the low-density EOS, and the low-density EOS is used at densities below the
intersection point. This matching method has the virtue of providing
monotonically increasing EOSs $p=p(\rho)$ without introducing additional
parameters.  The method accommodates a region of parameter space larger than
that spanned by the collection of candidate EOSs. It does, however, omit EOSs
with values of $p_1$ and $\Gamma_1$ that are incompatible, for which the slope
of the $\log p$ vs $\log \rho$ curve is too shallow to reach the pressure $p_1$
from the low-density part of the EOS.   
 
The accuracy with which a piecewise polytrope $\{\rho_i, K_i, \Gamma_i\}$,
approximates a candidate EOS is measured by the root mean square or rms
residual of the fit to $m$ tabulated points $\rho_j,p_j$:
\begin{eqnarray}
\sqrt{\frac{1}{m}\left(\sum_{i} \sum_{
\substack{ j \\ \rho_i < \rho_j \leq \rho_{i+1} }} 
\left(\log p_j - \log K_i + \Gamma_i \log \rho_j \right)^2 
\right)}
\label{eq:residual}
\end{eqnarray}
In each case, we compute the residual only up to the maximum density $\rho_{\rm
max}$ that can occur in a stable star -- the central density of the maximum
mass nonrotating model based on the candidate EOS.
Because astrophysical observations can,
in principle, depend on the high-density EOS only up to the value of $\rho_{\rm
max}$ for that EOS, only the accuracy of the fit below $\rho_{\rm max}$ for
each candidate EOS is relevant to our choice of parameter space.    

For a given parameterization, we find for each candidate EOS the smallest value
of the residual over the corresponding parameter space and the parameter values
for which it is a minimum.  In particular, the MINPACK nonlinear least-squares
routine LMDIF, based on the Levenberg-Marquardt algorithm, is used to minimize
the sum of squares of the difference between the logarithm of the
pressure-density points in the specified density range and the logarithm of the
piecewise polytrope formula, which is a linear fit in each region to the
$\log(p)[\log\rho]$ curve of the candidate EOS.  The nonlinear routine allows
the dividing points between regions to be varied. 

Even with a robust algorithm, the nonlinear fitting with varying dividing
densities is sensitive to initial conditions. Multiple initial parameters for
free fits are constructed using fixed-region fits of several possible dividing
densities, and the global minimum of the resulting residuals is taken to
indicate the best fit for the candidate EOS.

We begin with a single polytropic region in the core, specified by two
parameters: the index $\Gamma_1$ and a pressure $p_1$ at some fixed density.
Here, with a single polytrope, the choice of that density is arbitrary; for
more than one polytropic piece, we will for convenience take that density to be
the dividing density $\rho_1$ between the first two polytropic regions.
Changing the value of $p_1$ moves the polytropic $p(\rho)$ curve up or down,
keeping the logarithmic slope $\Gamma_1=d\log p/d\log\rho$ fixed. The
low-density SLy EOS is fixed, and the density $\rho_0$  where the polytropic
EOS intersects SLy changes as $p_1$ changes. The polytropic index $K_1$ is
determined by Eq.~(\ref{eq:Kcont}).  This is referred to as a one free piece
fit.  

We then consider two polytropic regions within the core, specified by the four
parameters $\{p_1, \Gamma_1, \rho_1, \Gamma_2\}$,  as well as three polytropic
regions specified by the six parameters $\{p_1, \Gamma_1, \rho_1, \Gamma_2,
\rho_3, \Gamma_3\}$, where, in each case, $p_1 \equiv p(\rho_1)$.  Again
changing $p_1$ translates the piecewise-polytropic EOS of the core up or down,
keeping its shape fixed.     While some EOSs are well approximated by a single
high-density polytrope, others require three pieces to capture the behaviour of
phase transitions at high density.

The six parameters required to specify three free polytropic pieces seems to
push the bounds of what may be reasonably constrained by a small set of
astrophysical measurements.  We find, however, that the collection of candidate
EOSs has a choice of dividing densities for which the residuals of the fit
exhibit a clear minimum.  This fact allows us to reduce the number of
parameters by fixing the densities that delimit the polytropic regions of the
piecewise polytrope.  A three fixed piece fit, using three polytropic regions
but with fixed $\rho_1 = 10^{14.7}$~g/cm$^3$ and $\rho_2 = 10^{15.0}$~g/cm$^3$,
is specified by the four parameters $p_1$, $\Gamma_1$, $\Gamma_2$, and
$\Gamma_3$.  The choice of $\rho_1$ and $\rho_2$ is discussed in Section
\ref{sec:fixed}.  Note that the density of departure from the fixed low-density
EOS is still a fitted parameter for this scheme.

\section{Best fits to Candidate EOS} 
\label{sec:fits} 
\subsection{Accuracy of alternative parameterizations}
\label{sec:accuracy}

The accuracy of each of the alternative choices of parameters discussed in the
last Section, measured by the rms residual of Eq.~(\ref{eq:residual}), is
portrayed in Table \ref{tab:FitResiduals}.  The Table lists the average and
maximum rms residuals over the set of 34 candidate EOSs.

\begin{table}[!htb]
\begin{center}
\caption[Average residuals from fitting candidate EOSs
]{
Average residuals resulting from fitting the set of candidate EOSs with various
types of piecewise polytropes. Free fits allow dividing densities between
pieces to vary. The fixed three piece fit uses $10^{14.7} $\,g/cm$^{3}$ or
roughly $1.85\rho_{\mathrm{nuc}}$ and $10^{15.0}$\,g/cm$^3$ or
$3.70\rho_{\mathrm{nuc}}$ for all EOSs. Tabled are the RMS residuals of the
best fits averaged over the set of candidates. The set of 34 candidates
includes 17 candidates containing only $npe\mu$ matter and 17 candidates with
hyperons, pion or kaon condensates, and/or quark matter. Fits are made to
tabled points in the high density region between $10^{14.3}$\,g/cm$^{3}$ or
$0.74\rho_{\mathrm{nuc}}$ and the central density of a maximum mass TOV star
calculated using that table\label{tab:FitResiduals}.}
\begin{tabular}{lccc}
\hline
\hline
Type of fit & All & $npe\mu$ & $K/\pi/h/q$\\
\hline
\multicolumn{4}{l}{Mean RMS residual}\\
\hline
One free piece&0.0386 & 0.0285& 0.0494\\
Two free pieces& 0.0147 & 0.0086& 0.0210\\
Three fixed pieces& 0.0127 & 0.0098& 0.0157 \\
Three free pieces& 0.0071 & 0.0056& 0.0086 \\
\hline
\hline
\multicolumn{4}{l}{Standard deviation of RMS residual}\\
\hline
One free piece&0.0213& 0.0161& 0.0209\\
Two free pieces&0.0150& 0.0060& 0.0188\\
Three fixed pieces& 0.0106 & 0.0063& 0.0130 \\
Three free pieces& 0.0081& 0.0039& 0.0107\\
\hline
\hline
\end{tabular}
\end{center}
\end{table}

For nucleon EOSs, the four-parameter fit of two free polytropic pieces models
the behaviour of candidates well; but this kind of four-parameter EOS does not
accurately fit EOSs with hyperons, kaon or pion condensates, and/or quark
matter. Many require three polytropic pieces to capture the stiffening around
nuclear density, a subsequent softer phase transition, and then final
stiffening. On the other hand, the six parameters required to specify three
free polytropic pieces exceeds  the bounds of what may be reasonably
constrained by the small set of model-independent astrophysical measurements.
As mentioned in the introduction, however, an alternative four parameter fit
can be made to all EOSs if the transition densities are held fixed for all
candidate EOSs. The choice of fixed transition densities, and the advantages of
such a parameterization over the two free piece fit, are discussed in the next
subsection.

The hybrid quark EOS ALF3, which incorporates a QCD correction parameter
for quark interactions, exhibits the worst-fit to a one-piece polytropic
EOS with residual 0.111, to the three-piece
fixed region EOS  with residual 0.042, and to
the three-piece varying region EOS  with
residual 0.042.
It has a residual from the two-piece fit of 0.044, somewhat less than the
worst fit EOS, BGN1H1, an effective-potential EOS that includes all
possible hyperons and has a two-piece fit residual of 0.056.

\subsection{Fixed region fit}
\label{sec:fixed}

\begin{figure}[!htb]
\includegraphics[width=3.4in]{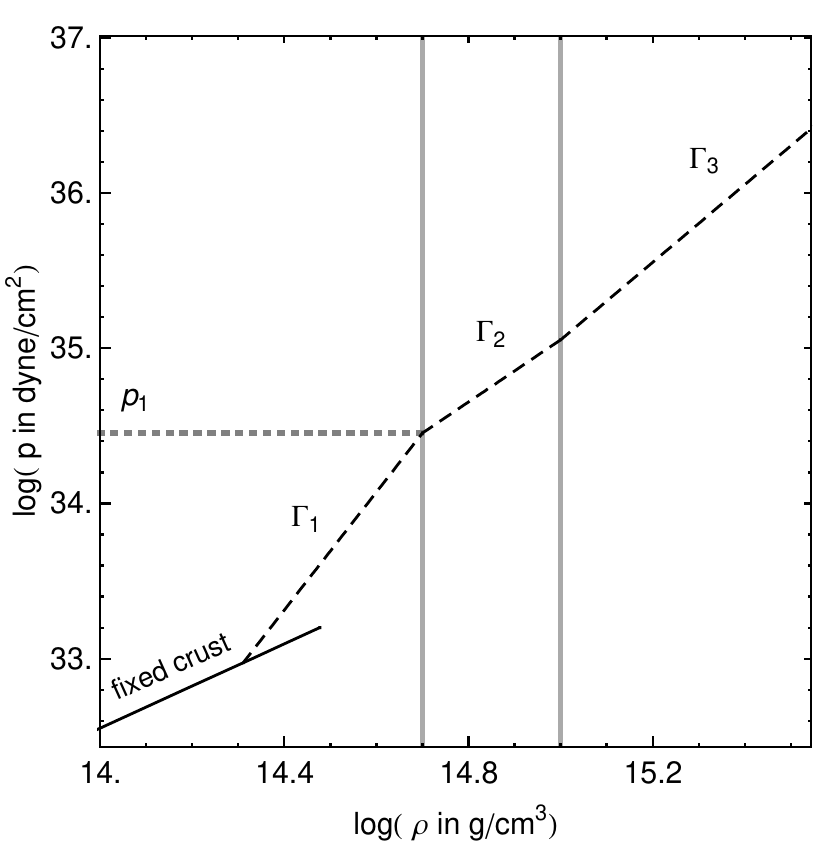}
\caption{ 
The fixed-region fit is parameterized by adiabatic indices $\{\Gamma_1,
\Gamma_2, \Gamma_3\}$ and by the pressure $p_1$ at the first dividing density.}
\label{fig:schematic} 
\end{figure}

A good fit with a minimal number of parameters is found for three regions with
a division between the first and second pieces fixed at $\rho_1 =
10^{14.7}$~g/cm$^3 = 1.85 \rho_{\mathrm{nuc}}$ and a division between the
second and third pieces fixed at $\rho_2 = 10^{15.0}$~g/cm$^3$. The EOS is
specified by choosing the adiabatic indices $\{\Gamma_1, \Gamma_2, \Gamma_3\}$
in each region, and the pressure $p_1$ at the first dividing density, $p_1 =
p_1(\rho_1)$.  A diagram of this parameterization is shown in
Fig.~\ref{fig:schematic}.
For this 4-parameter EOS, best fit parameters for each candidate EOS give a
residual of $0.043$ or better, with the average residual over 34 candidate EOSs
of $0.013$.

\begin{figure}[!hbt]
\includegraphics[width=3.4in]{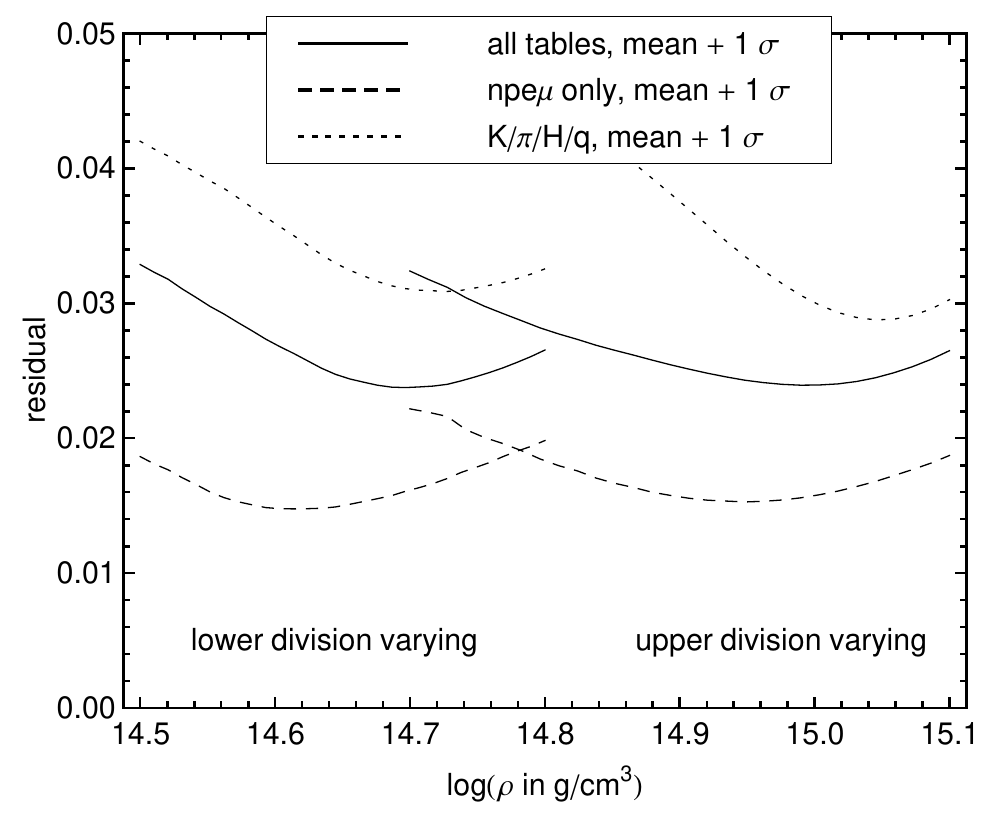}\\
\caption[Optimal dividing densities for subsets of candidate EOS]{ Subsets of
EOSs with and without kaons, hyperons, meson condensates, or quarks, show a
fairly robust choice of dividing densities whose fit to the candidate EOSs
minimizes residual error.  The mean plus one standard deviation of residuals
for each subset of candidate EOSs is plotted against the choice of lower and
upper dividing densities $\rho_1$ and $\rho_2$.  The left curves show mean
residual versus $\rho_1$ with $\rho_2$ fixed at $10^{15.0}$\,g/cm$^3$. The
right three curves show mean residual versus $\rho_2$, with $\rho_1$ fixed at
$10^{14.7}$\,g/cm$^3$.}
\label{fig:divsets}
\end{figure}

The dividing densities for our parameterized EOS were chosen by minimizing the
rms residuals over the set of 34 candidate EOSs.  For two dividing densities,
this is a two-dimensional minimization problem, which was solved by alternating
between minimizing average rms residual for upper or lower density while
holding the other density fixed.  The location of the best dividing points is
fairly robust over the subclasses of EOSs, as illustrated in
Fig.~\ref{fig:divsets}.  

% For fixed crust EOS, the parameter $\rho_1$ can be reinterpreted as
% specifying an overall pressure shift for an equation of state with given
% $\Gamma_1$, $\Gamma_2$, and $\Gamma_3$. Instead of $\rho_1$, pressure at any
% given density can be used to specify the fixed region piecewise polytrope in
% the core.

With the dividing points fixed, taking the pressure $p_1$ to be the pressure at
$\rho_1=1.85 \rho_{\mathrm{nuc}}$, is indicated by empirical work of Lattimer
and Prakash \cite{Lattimer:2000nx} that finds a strong correlation between
pressure at fixed density (near this value) and the radius of $1.4 M_\odot$
neutron stars.  This choice of parameter allows us to examine (in
Sec.~\ref{sec:moment}) the relation between $p_1$ and the radius; and we
expect a similar correlation between $p_1$ and the frequency at which
neutron-star inspiral dramatically departs from point-particle inspiral for
neutron stars near this mass.    

The following considerations dictate our choice of the four-parameter space
associated with three polytropic pieces with two fixed dividing densities.
First, as mentioned above, we regard the additional two parameters needed for
three free pieces as too great a price to pay for the moderate increase in
accuracy.  The comparison, then, is between two four-parameter spaces:
polytropes with two free pieces and polytropes with three pieces and fixed
dividing densities.  

Here there are two key differences.  Observations of pulsars that are not
accreting  indicate
masses below 1.45 $M_\odot$ (see
Sec.~\ref{sec:constraints}), and the
central density of these stars is below $\rho_2$ for almost all EOSs. Then only
the three parameters $\{p_1 , \Gamma_1 , \Gamma_2 \}$ of the fixed piece
parameterization are required to specify the EOS for moderate mass neutron
stars. This class of observations can then be treated as a set of constraints
on a 3-dimensional parameter space. Similarly, because maximum-mass neutron
stars ordinarily have most matter in regions with densities greater than the
first dividing density, their structure is insensitive to the first adiabatic
index.  The three piece parameterization does a significantly better job above
$\rho_2$ because phase transitions above that density require a third
polytropic index $\Gamma_3$. If the remaining three parameters can be
determined by pulsar observations, then observations of more massive, accreting
stars can constrain $\Gamma_3$.  

The best fit parameter values are shown in Fig.~\ref{fig:candidatefits} and
listed in Table~\ref{tab:constraints} of Appendix~\ref{ap:paramacc}.  The worst fits
of the fixed region fit are the hybrid quark EOSs ALF1 and
ALF2, and the hyperon-incorporating EOS BGN1H1. For
BGN1H1, the relatively large residual is due to the fact that the
best fit dividing densities of BGN1H1 differ strongly from the
average best dividing densities. Although BGN1H1 is well fit by
three pieces with floating densities, the reduction to a four-parameter fit
limits the resolution of EOSs with such
structure.  The hybrid quark EOSs,
however, have more complex structure that is difficult to resolve accurately
with a small number of polytropic pieces.  Still, the best-fit polytrope EOS is
able to reproduce the neutron star properties predicted by the hybrid quark
EOS.
\begin{figure}[!hbt]
\begin{center}
\includegraphics[width=3.4in]{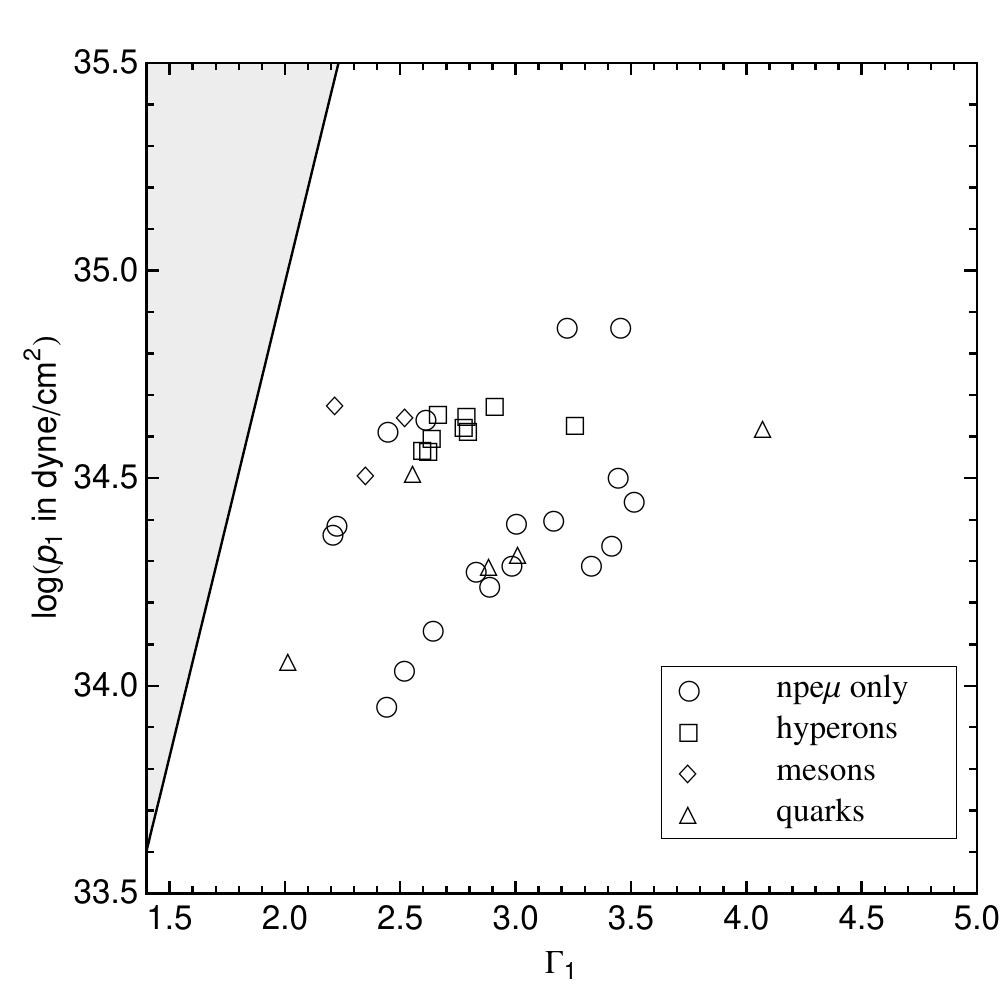}
\includegraphics[width=3.4in]{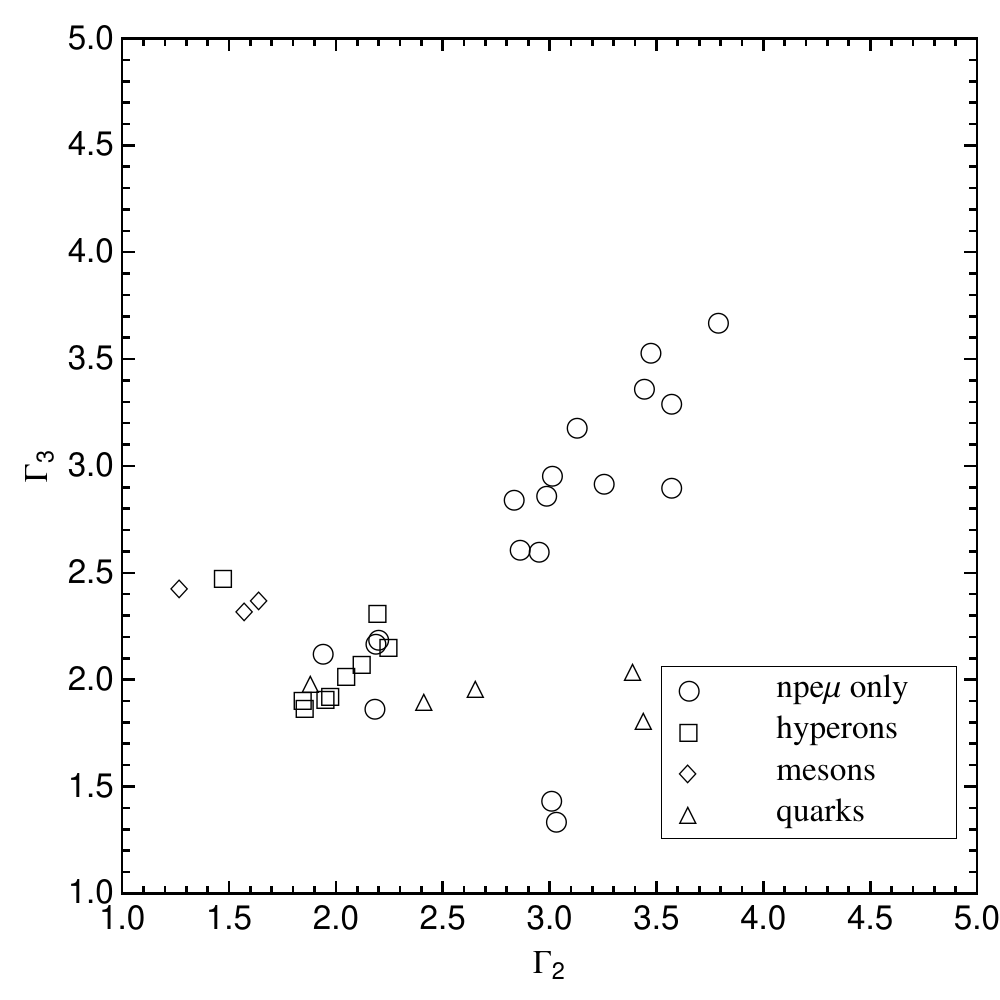}
\end{center}
\caption{
\label{fig:candidatefits}
Parameterized EOS fits to the set of 34 candidate EOS tables.  There are 17
EOSs with only ordinary nuclear matter (n,p,e,$\mu$);  9 have only hyperons in
addition to ordinary matter;  3 include meson condensates plus ordinary matter;
5 include quarks plus other matter (PCL2 also has hyperons).
$\Gamma_2<3.5$ and $\Gamma_3<2.5$ for all EOSs with hyperons, meson
condensates, and/or quark cores. The shaded region
corresponds to incompatible values of $p_1$ and $\Gamma_1$, 
as discussed in the text.}
\end{figure}

In Appendix~\ref{ap:paramacc}, Table~\ref{tab:constraints} also shows the various
neutron star structure characteristics for each EOS compared to the values of
the best-fit piecewise polytrope parameterization for the core.  The mean error
and standard deviation for each characteristic is also listed.

\section{Astrophysical constraints on the parameter space}
\label{sec:constraints} 

Adopting a parameterized EOS allows one to phrase each observational constraint
as a restriction to a subset of the parameter space.  In subsections A--D we
find the constraints imposed by causality, by the maximum observed neutron-star
mass and the maximum observed neutron-star spin, and by a possible observation
of gravitational redshift.  We then examine, in subsection E, constraints
from the simultaneous measurement of mass and moment of inertia and of mass
and radius. We exhibit in subsection F the combined constraint imposed by
causality, maximum observed mass, and a future moment-of-inertia
measurement of a star with known mass.

In exhibiting the constraints, we show a region of the 4-dimensional parameter
space larger than that allowed by the presumed uncertainty in the EOS -- large
enough, in particular, to encompass the 34 candidate EOSs considered above.
The graphs in Fig.~\ref{fig:candidatefits} display the ranges
$10^{33.5}\mbox{dyne/cm}^2<p_1<10^{35.5}\mbox{dyne/cm}^2$, $1.4<\Gamma_1<5.0$,
$1.0<\Gamma_2<5.0$, and $1.0<\Gamma_3<5.0$.  Also shown is the location in
parameter space of each candidate EOS, defined as the set of parameters that
provide the best fit to that EOS.  The shaded region in the top graph
corresponds to incompatible values of $p_1$ and $\Gamma_1$ mentioned in
Sect.~\ref{sec:fit-method}: These are values for which the pressure $p_1$ is so
large and $\Gamma_1$ so small that no curve $p(\rho)$ can start from the
low-density EOS above neutron drip and reach $p_1$ at $\rho_1$ unless the
slope $d\log p/d\log \rho$ exceeds $\Gamma_1$.

To find the constraints on the parameterized EOS imposed by the maximum
observed mass and spin, one finds the maximum mass and spin of stable neutron
stars based on the EOS associated with each point of parameter space.  A
subtlety in determining these maximum values arises from a break in the
sequence of stable equilibria -- an island of unstable configurations -- for
some EOSs.  The unstable island is typically associated with phase transitions
in a way we now describe. 

Spherical Newtonian stars described by EOSs of the form $p=p(\rho)$ are unstable
when an average value $\bar\Gamma$ of the adiabatic index falls below 4/3.  The
stronger-than-Newtonian gravity of relativistic stars means that instability
sets in for larger values of $\bar\Gamma$, and it is ordinarily this increasing
strength of gravity that sets an upper limit on neutron-star mass.  EOSs with
phase transitions, however, temporarily soften above the critical density and
then stiffen again at higher densities.  As a result, configurations whose
inner core has density just above the critical density can be unstable, while
configurations with greater central density can again be stable. Models with
this behavior are considered, for example, by Glendenning and
Kettner~\cite{Glendenning:1998ag}, Bejger et
al.~\cite{BejgerHaenselZdunik2005phase} and by Zdunik et
al.~\cite{ZdunikBejgerHaensel2006phase} (these latter authors, in fact, use
piecewise polytropic EOSs to model phase transitions).

\begin{figure}[!hbt]
\begin{center}
\includegraphics[width=3.4in]{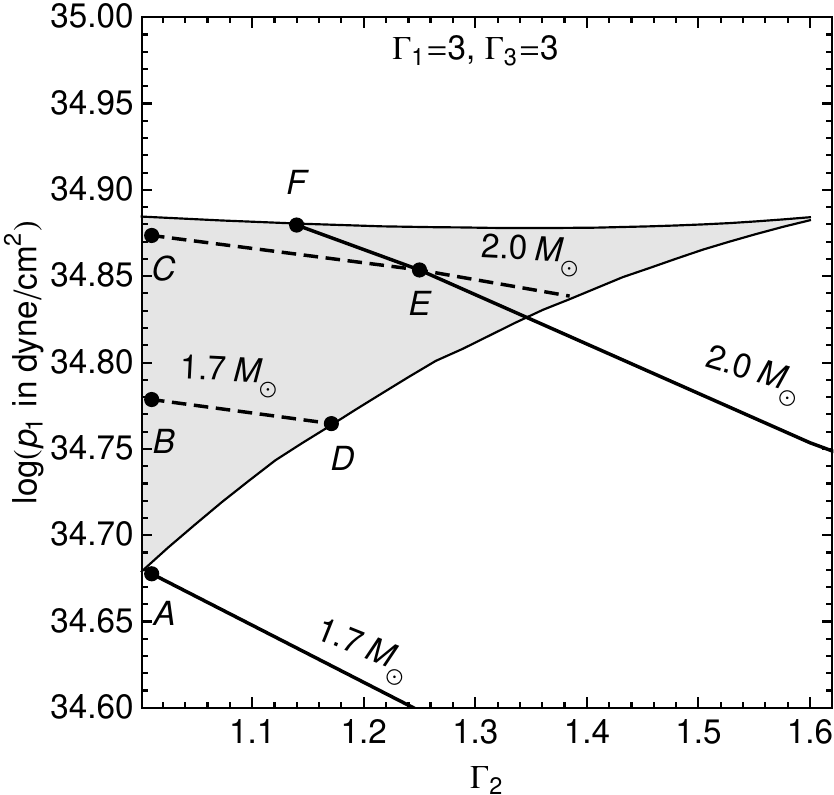}
\includegraphics[width=3.4in]{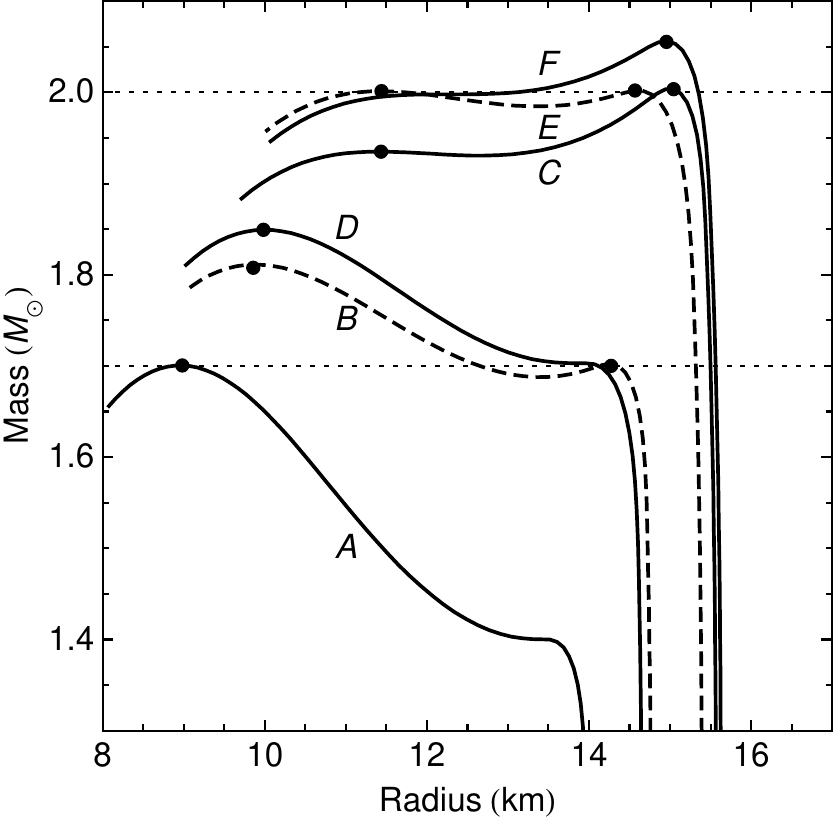}
\end{center}
\caption{
\label{fig:2branches}
The region in parameter space where two stable neutron-star sequences can occur
is shaded in the top figure.  Contours of constant maximum mass are also shown.
The higher central density maximum mass contour is solid while the lower
central density maximum mass contour is dashed.  Mass-radius curves are plotted
for several EOSs in the bottom figure.  Although difficult to see, EOS C does
in fact have a second stable sequence.} 
\end{figure}

For our parameterized EOS, instability islands of this kind can occur for
$\Gamma_2\lesssim 2$, when $\Gamma_1\gtrsim 2$ and $\Gamma_3\gtrsim 2$.  A
slice of the four-dimensional parameter space with constant $\Gamma_1$ and
$\Gamma_3$ is displayed in Fig.~\ref{fig:2branches}.  The shaded region
corresponds to EOSs with islands of instability.  Contours are also shown for
which the maximum mass for each EOS has the constant value $1.7 M_\odot$ (lower
contour) and $2.0 M_\odot$ (upper contour).  

An instability point along a sequence of stellar models with constant angular
momentum occurs when the mass is maximum.  On a mass-radius curve, stability is
lost in the direction for which the curve turns counterclockwise about the
maximum mass, regained when it turns clockwise.  In the bottom graph of
Fig.~\ref{fig:2branches}, mass-radius curves are plotted for six EOSs, labeled
A-F, associated with  six correspondingly labeled EOSs in the top figure. The
sequences associated with EOSs B, C and E have two maximum masses (marked by
black dots in the lower figure) separated by a minimum mass.  As one moves
along the sequence from larger to smaller radius -- from lower to higher
density, stability is temporarily lost at the first maximum mass, regained at
the minimum mass, and permanently lost at the second maximum mass.  

It is clear from each graph in Fig.~\ref{fig:2branches} that either of the two
local maxima of mass can be the global maximum.  On the lower boundary
(containing EOSs A and D), the lower density maximum mass first appears, but
the upper-density maximum remains the global maximum in a neighborhood of the
boundary.  Above the upper boundary (containing EOS F), the higher-density
maximum has disappeared, and near the upper boundary the lower-density maximum
is the global maximum.  

\subsection{Causality}

For an EOS to be considered physically reasonable, the adiabatic speed of sound
cannot exceed the speed of light. Perfect fluids have causal time evolutions
(satisfy hyperbolic equations with characteristics within the light cone) only
if $v_{\rm s}$ (the phase velocity of sound) is less than the speed of light.  An EOS
is ruled out by causality if $v_{\rm s}>1$ for densities below the central density
$\rho_{\rm max}$ of the maximum-mass neutron star for that EOS.  An EOS that
becomes acausal beyond $\rho_{\rm max}$ at density higher than this can always
be altered for $\rho>\rho_{\rm max}$ to a causal EOS. Because the original and
altered EOS yield identical sequences of neutron stars, causality should not be
used to rule out parameters that give formally acausal EOSs above $\rho_{\rm
max}$. 

\begin{figure}[!htb]
\begin{center}
\includegraphics[width=3.4in]{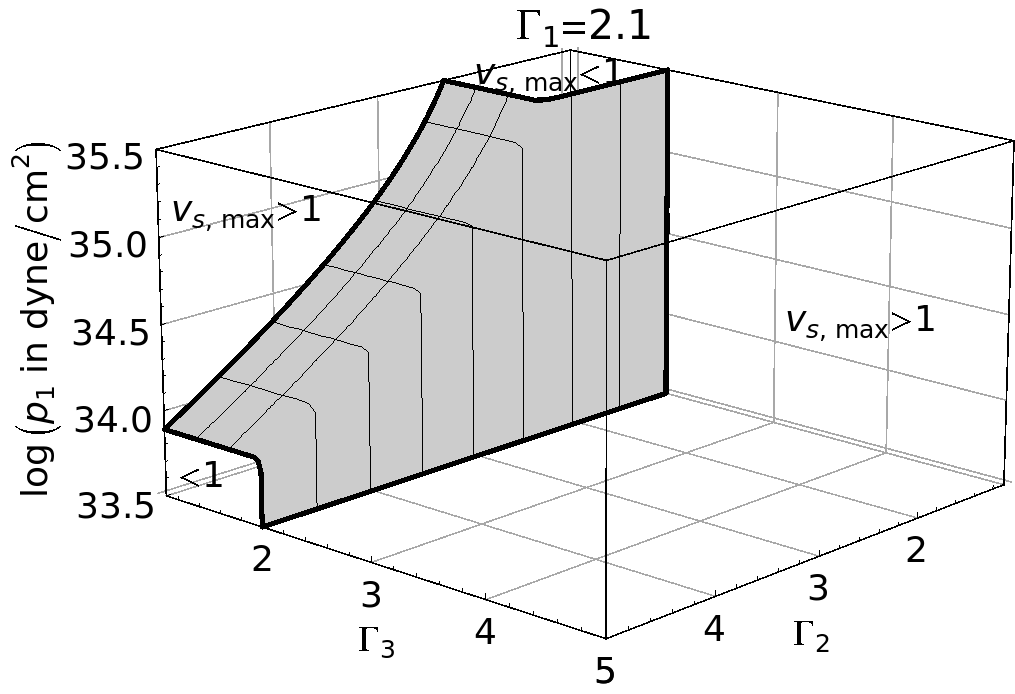}
\includegraphics[width=3.4in]{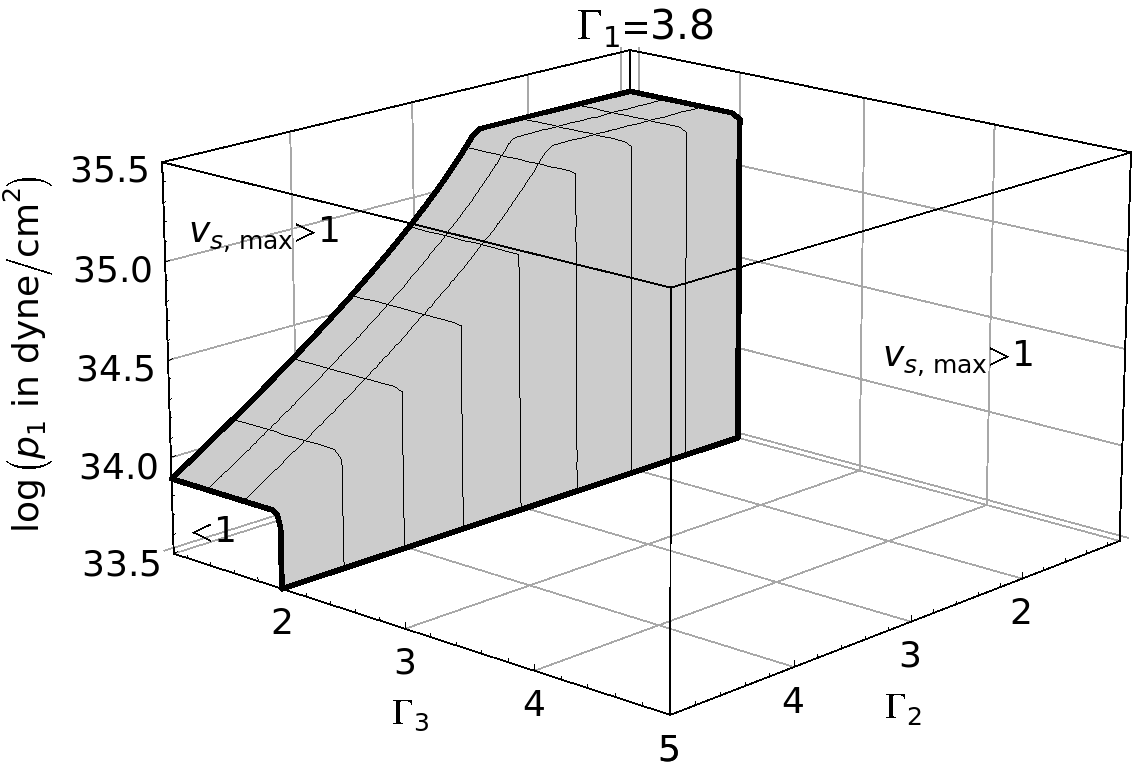}
\caption{Causality constraints are shown for two values of $\Gamma_1$.  For
each EOS in the parameter space the maximum speed of sound over all densities
is used.  The shaded surface separates the EOS parameter space into a region
behind the surface allowed by causality (labeled $v_{\rm s,max}<1$) and a region in which
corresponding EOSs violate causality at any density (labeled $v_{\rm s,max}>1$).}
\label{fig:vsound1}
\end{center}
\end{figure}

We exhibit the causality constraint in two ways, first by simply requiring that
each piecewise polytrope be causal at all densities and then by requiring only
that it be causal below $\rho_{\rm max}$.  The first, unphysically strong,
constraint, shown in Fig.~\ref{fig:vsound1}, is useful for an intuitive
understanding of the constraint: The speed of sound is a measure of the
stiffness of the EOS, and requiring causality eliminates the largest values of
$\Gamma_i$ and $p_1$.  

\begin{figure}[!htb]
\begin{center}
\includegraphics[width=3.4in]{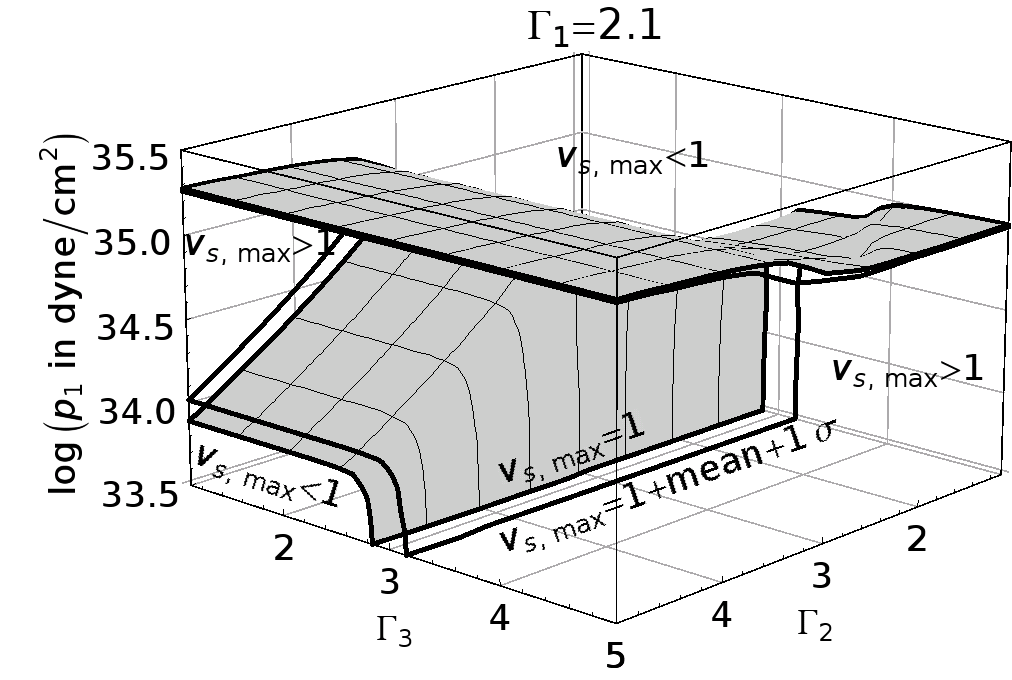}
\includegraphics[width=3.4in]{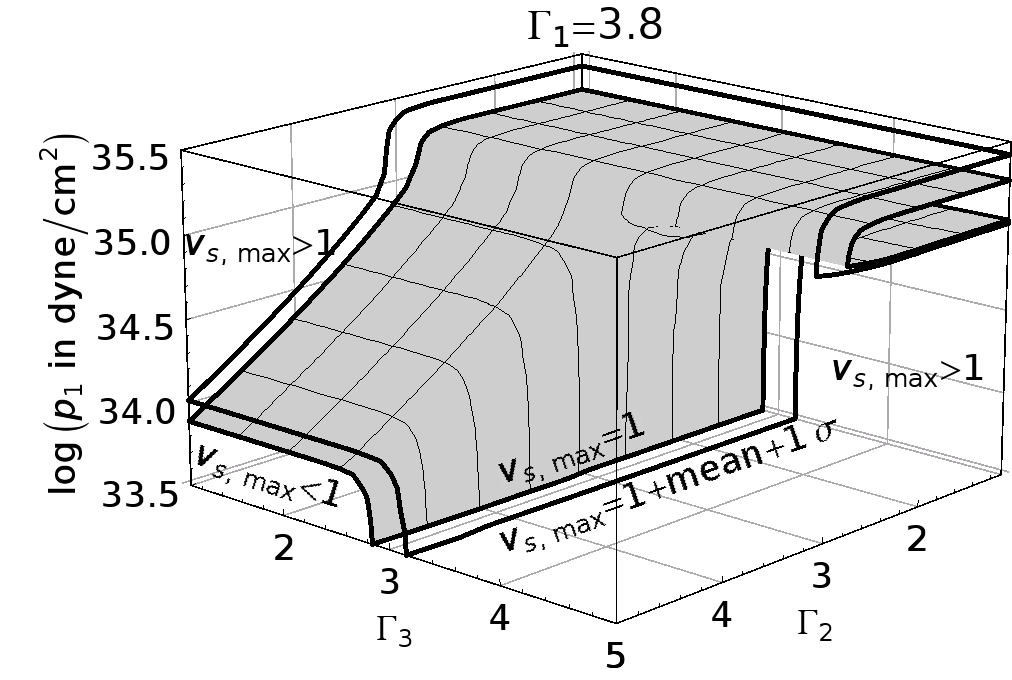}
\caption{Causality constraint as in  Fig.~\ref{fig:vsound1}.  However,
here, only the maximum speed of sound up to the central density of the maximum
mass star is considered.  A
second, outlined surface shows a weaker constraint to accommodate the expected
error in the speed of sound associated with a piecewise polytropic
approximation to an EOS.  With $\sigma$ the standard deviation in
$v_{\rm s,max}$ between
an EOS and its parameterized representation, as measured by the collection of
candidate EOSs, the outlined surface depicts $v_{\rm s,max} =1+$mean$+1\sigma=1.12$
constraint.}
\label{fig:vsound}
\end{center}
\end{figure}

Fig.~\ref{fig:vsound} shows the result of restricting the constraint to
densities below $\rho_{\rm max}$, with the speed of sound given by
Eq.~(\ref{eq:vs}).  A second surface is shown to account for the inaccuracy
with which a piecewise polytropic approximation to an EOS represents the speed
of sound.  In all but one case (BGN1H1) the fits to the candidate EOSs
overpredict the maximum speed of sound, but none of the fits to the candidate
EOSs mispredict whether the candidate EOS is causal or acausal by more than
11\% (fractional difference between fit and candidate).  We adopt as a suitable
causality constraint a restriction to a region bounded by the surface $v_{\rm
s, max}=1+$mean$+1\sigma=1.12$, corresponding to the mean plus one standard
deviation in the error between $v_{\rm s, max}$ for the candidate and best fit
EOSs.  

In the lower parts of each graph in Fig.~\ref{fig:vsound}, where
$p_1<10^{35}$~dyne/cm$^2$, the bounding surface has the character of the first
causality constraint, with the restriction on each of the three variables $p_1,
\Gamma_2$ and $\Gamma_3$ becoming more stringent as the other parameters
increase, and with $\Gamma_3$ restricted to be less than about $3$.  In this
low-pressure part of each graph, the surface is almost completely independent
of the value of $\Gamma_1$:
Because the constraint takes the form $\Gamma_1 p / (\epsilon+p) \le c^2$ (for
$p\ll\epsilon$) and $p<p_1$ is so low, the constraint rules out values of
$\Gamma_1$ only at or beyond the maximum $\Gamma_1$ we consider.

In the upper part of each graph, where $p_1>10^{35}$~dyne/cm$^2$,
unexpected features arise from the fact that we  impose the causality
constraint only below the maximum density of stable neutron stars -- below the
central density of the maximum-mass star.    

The most striking feature is the way the constraint surface turns over in the
upper part of the top graph, where $p_1>10^{35}$~dyne/cm$^2$, in a way that
allows arbitrarily large values of $p_1$.  This occurs because, when $p_1$ is
large, the density of the maximum-mass star is small, and a violation of
causality typically requires high density.  That is, when the density is low, the ratio
$p/(\epsilon+p)$ in Eq.~(\ref{eq:vs}) is
small.  As a result, in the top graph, $v_{\rm s}$ remains too small to violate
causality before the maximum density is reached.  In the bottom graph, with
$\Gamma_1=3.8$, $\Gamma_1$ is now large enough in Eq.~(\ref{eq:vs}) that the EOS becomes acausal just below the transition to
$\Gamma_2$.  This is the same effect that places the upper limit on $p_1$
seen in the second graph of Fig.~\ref{fig:vsound1}.

A second feature of the upper parts of each graph is the exact independence of
the bounding surface on $\Gamma_3$. The reason is simply that in this part of
the parameter space the central density of the maximum mass star is below
$\rho_2$, implying that no stable neutron stars see $\Gamma_3$.

Finally, we note that in both graphs, for small $\Gamma_2$ (the right of the graph), the EOSs yield the sequences
mentioned above, in which an island of instability separates two stable
sequences, each ending at a local maximum of the mass. Requiring $v_{\rm
s,max}$ to satisfy causality for both stable regions rules out EOSs below
the lower part of the bifurcated surface.  

\subsection{Maximum Mass}

\begin{figure}[!htb]
\begin{center}
\includegraphics[width=3.4in]{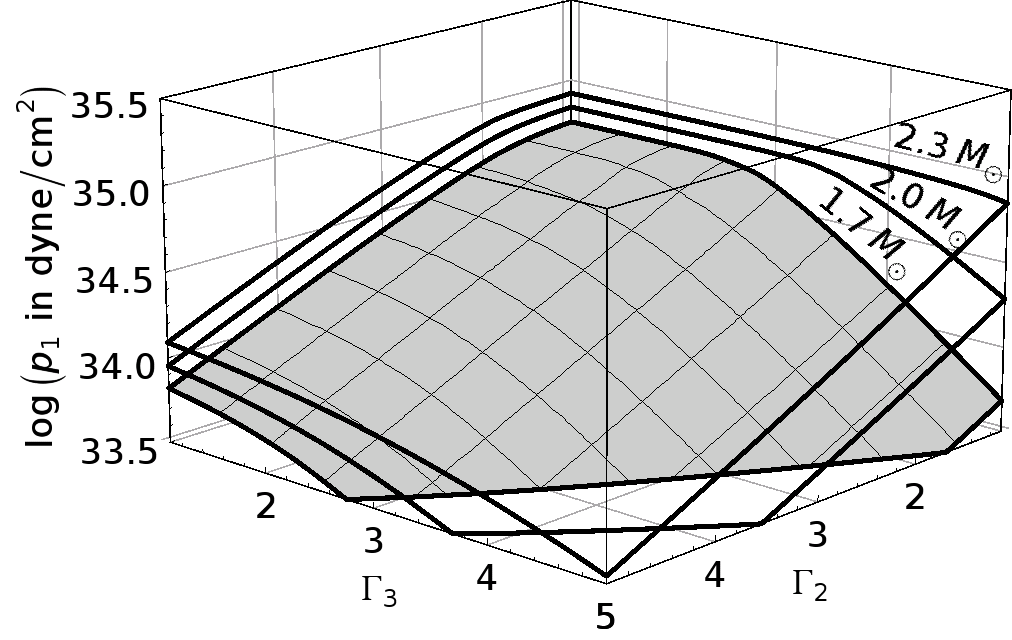}
\caption{The above surfaces represent the set of parameters that result in a
constant maximum mass.  An observation of a massive neutron
star constrains the equation of state to lie above the corresponding surface.
$\Gamma_1$ is set to the least constraining value at each point.  The lower
shaded surface represents $M_{\rm max}=1.7~M_\odot$; the middle and upper
(outlined) surfaces represent $M_{\rm max}=2.0~M_\odot$ and $M_{\rm
max}=2.3~M_\odot$ respectively.}
\label{fig:mall}
\end{center}
\end{figure}

A stringent observational constraint on the EOS parameter space is set by
the largest observed neutron-star mass.
Unfortunately, the highest claimed masses are also subject to the highest
uncertainties and systematic errors. The most reliable measurements come from
observations of radio pulsars in binaries with neutron star companions.
The masses with tightest error bars (about 0.01~$M_\odot$) cluster
about 1.4~$M_\odot$~\cite{Lattimer:2006xb}. Recent observations of
millisecond
pulsars in globular clusters with non-neutron star companions have yielded
higher masses:
Ter~5I and Ter~5J~\cite{Ransom:2005ae}, M5B~\cite{Freire:2007xg},
PSR~J1903+0327~\cite{Champion:2008ge}, and
PSR~J0437-4715~\cite{Verbiest:2008gy} all have 95\% confidence limits of
about 1.7~$M_\odot$, and the corresponding limit for
NGC~6440B~\cite{Freire:2007jd} is about 2.3~$M_\odot$.
However these systems are
more prone to systematic errors: The pulsar mass is obtained by assuming that
the periastron advance of the orbit is due to general relativity.  Periastron
advance can also arise from rotational deformation of the companion, which is
negligible for a neutron star but could be much greater for pulsars
which have white dwarf or main sequence star companions.
Also the mass measurement is affected by inclination angle, which is known
only for the very nearby PSR~J0437-4715.
And with the accumulation of observations of these eccentric binary systems
(now about a dozen) it becomes more likely that the anomalously high figure
for NGC~6440B is a statistical fluke.
Fig.~\ref{fig:mall} shows the
constraint on the EOS placed by the existence of 1.7~$M_\odot$ neutron stars,
which we regard as secure. Also shown in the figure are the surfaces associated
with maximum masses  of 2.0~$M_\odot$ and 2.3~$M_\odot$.

Since all of the candidate high-mass pulsars are spinning slowly enough that
the rotational contribution to their structure is negligible, the constraint
associated with their observed masses can be obtained by computing the maximum
mass of nonrotating neutron stars.  Corresponding to  each point in the
parameter space is a sequence of neutron stars based on the associated
parameterized EOS; and a point of parameter space is ruled out if the
corresponding sequence has maximum mass below the largest observed mass.  We
exhibit here the division of parameter space into regions allowed and forbidden
by given values of the largest observed mass.  

We plot contours of constant maximum mass in Fig.~\ref{fig:mall}.   Because
EOSs below a maximum mass contour produce stars with lower maximum masses, the
parameter space below these surfaces is ruled out.  The error in the maximum
mass between the candidate and best fit piecewise polytropic
EOSs is $|$mean$|+1\sigma=1.7\%$ (the magnitude of the mean error plus one
standard deviation in the error over the 34 candidate EOSs), so the parameters
that best fit the true EOS are unlikely to be below this surface.

The surfaces of Fig.~\ref{fig:mall} have minimal dependence on $\Gamma_1$,
indicating that the maximum mass is determined primarily by features of the EOS
above $\rho_1$. In Fig.~\ref{fig:mall} we have set $\Gamma_1$ to the least
constraining value in the range we consider -- to the value that gives the
largest maximum mass at each point in $\{p_1,\Gamma_2,\Gamma_3\}$ space.
Varying $\Gamma_1$ causes the contours to shift up, constraining the parameter
space further, by a maximum of $10^{0.2}$~dyne/cm$^2$.  The dependence of the
contour on $\Gamma_1$ is most significant for large values of $p_1$ where the
average density of a star is lower.  The dependence on $\Gamma_1$ decreases significantly as
$p_1$ decreases.

As discussed above, some of the EOSs produce sequences of spherical neutron
stars with an island of instability separating two stable sequences, each with
a local maximum of the mass. As shown in Fig.~\ref{fig:2branches}, this causes
a contour in parameter space of constant maximum mass to split into two
surfaces, one surface of parameters which has this maximum mass at the lower
$\rho_{\rm c}$ local maximum and another surface of parameters which has this maximum
mass at higher $\rho_{\rm c}$ branches.  Since such EOSs allow stable models up to
the largest of their local maxima, we use the least constraining surface
(representing the global maximum mass) when ruling out points in parameter
space. 

\subsection{Gravitational redshift}

  We turn next to the constraint set by an observed redshift of spectral lines
from the surface of a neutron star.  We consider here only stars for which the
broadening due to rotation is negligible and restrict our discussion to
spherical models.  The redshift is then $z=(1-2M/R)^{-1/2}-1$, and measuring it
is equivalent to measuring the ratio $M/R$.  With no independent measurement of
mass or radius, the associated constraint again restricts the parameter space
to one side of a surface, to the EOSs that allow a redshift as large as the
largest observed shift. \footnote{ One could also imagine a measured redshift
small enough to rule out a class of EOSs.  The minimum redshift for each EOS,
however, occurs for a star whose central density is below nuclear density.  Its
value, $z\approx 5\times 10^{-4}$, thus depends only on the EOS below nuclear
density. (See, for example Haensel et al.\cite{hzd02}.)} For spherical models,
the configuration with maximum redshift for a given EOS is ordinarily the
maximum-mass star. By increasing $p_1,\Gamma_2$ or $\Gamma_3$, one stiffens the
core, increasing the maximum mass, but also increasing the radius at fixed
mass.  The outcome of the competition usually, but not always, yields increased
redshift for larger values of these three parameters; that is, the increased
maximum mass dominates the effect of increased radius at fixed mass for all but
the largest values of $p_1$.   
  
Cottam, Paerels, and Mendez~\cite{Cottam:2002cu} claim to have observed
spectral lines from EXO~0748-676 with a gravitational redshift of $z=0.35$.
With three spectral lines agreeing on the
redshift, the identification of the spectral features with iron lines is better
founded than other claims involving only a single line.  The identification
remains in doubt, however, because the claimed lines have not been seen in
subsequent bursts~\cite{Cottam:2007cd}.  There is also a claim of a
simultaneous mass-radius measurement of this system using Eddington-limited
photospheric expansion x-ray bursts~\cite{Ozel:2006bv} which would rule out
many EOSs.  This claim is controversial, because the 95\% confidence interval
is too wide to rule out much of the parameter space, and we believe the
potential for systematic error is understated.  However, the gravitational
redshift is consistent with the earlier claim of 0.35.  Thus we treat $z=0.35$
as a tentative constraint.  We also exhibit the constraint that would be
associated with a measurement of $z=0.45$.  

\begin{figure}[!htb]
\begin{center}
\includegraphics[width=3.4in]{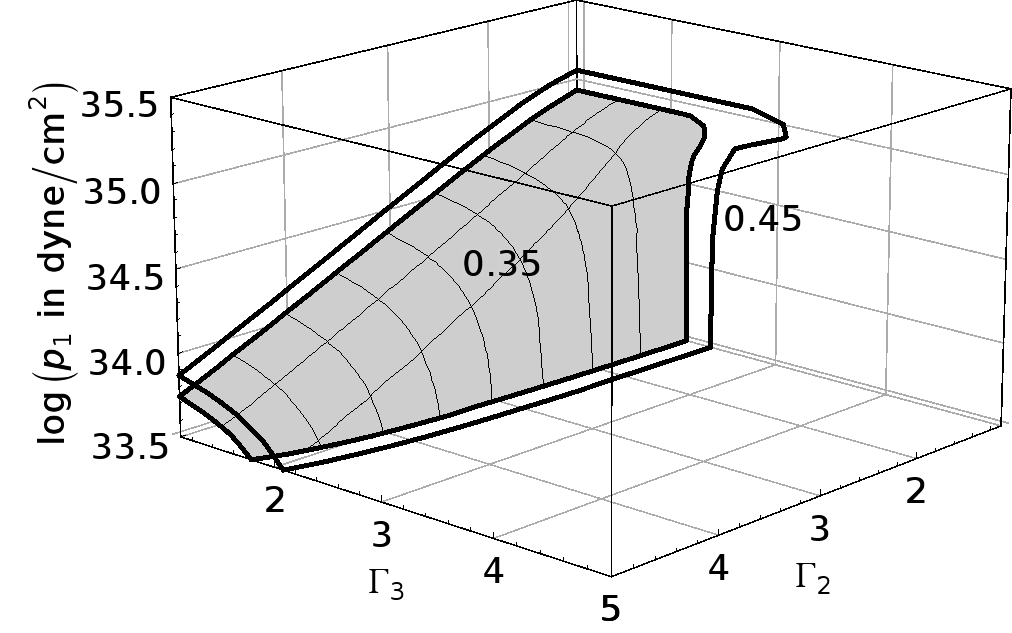}
\caption{Surfaces in the EOS parameter space for which the maximum redshift of
stable spherical neutron stars has the values 0.35 (shaded surface) and 0.45
(outlined surface).  A measured redshift from the surface of a neutron star
would exclude the region of parameter space behind the corresponding surface.
$\Gamma_1$ is fixed at 5.0, the least constraining in the range we considered.}
\label{fig:zall}
\end{center}
\end{figure}

Our parameterization can reproduce the maximum redshift of tabulated EOSs to
3.2\% (mean+1$\sigma$).  Figure~\ref{fig:zall} displays surfaces of constant
redshift $z=0.35$ and $z=0.45$ for the least constraining value of $\Gamma_1=5$
in the range we consider.  Surfaces with different values of $\Gamma_1$ are
virtually identical for $p_1<10^{34.8}$~dyne/cm$^2$, but diverge for higher
pressures when $\Gamma_1$ is small ($\lesssim 2.5$).  In the displayed
parameter space, points in front of the $z=0.35$ surface, corresponding to
stiffer EOSs in the inner core, are allowed by the potential $z=0.35$
measurement.  From the location of the $z=0.35$ and $z=0.45$ surfaces, it is
clear that, without an upper limit on $\Gamma_1 \lesssim 2.5$, an observed
redshift significantly higher than $0.35$ is needed to constrain the parameter
space.  In particular, most of the parameter space ruled out by $z=0.35$ is
already ruled out by the $M_{\rm max}=1.7~M_\odot$ constraint displayed in
Fig.~\ref{fig:mall}.

\subsection{Maximum Spin}

Observations of rapidly rotating neutron stars can also constrain the EOS.  The
highest uncontroversial spin frequency is observed in pulsar Ter~5AD at
716~Hz~\cite{Hessels:2006ze}.  There is a claim of 1122~Hz inferred from
oscillations in x-ray bursts from XTE~J1239-285~\cite{Kaaret:2006gr}, but this
is controversial because the statistical significance is relatively low, the
signal could be contaminated by the details of the burst mechanism such as
fallback of burning material, and the observation has not been repeated.

The maximum angular velocity of a uniformly rotating star occurs at the Kepler
or mass-shedding limit, $\Omega_{\rm K}$, with the star rotating at the speed of a
satellite in circular orbit at the equator.  For a given EOS, the configuration
with maximum spin is the stable configuration with highest central density
along the sequence of stars rotating at their Kepler limit.  An EOS thus
maximizes rotation if it maximizes the gravitational force at the equator of a
rotating star -- if it allows stars of large mass and small radius. To allow
high mass stars, the EOS must be stiff at high density, and for the radius of
the high-mass configuration to be small, the EOS must be softer at low density,
allowing greater compression in the outer part of the star
\cite{skf97,glendenning92}. In our parameter space, a high angular velocity
then restricts one to a region with large values of $\Gamma_2$ and $\Gamma_3$,
and small values of $p_1$ and $\Gamma_1$.  

As with the maximum mass, the maximum frequency is most sensitive to the
parameter $p_1$, but the frequency constraint complements the maximum mass
constraint by placing an upper limit on $p_1$ over the parameter space, rather
than  a lower limit. 

To calculate the maximum rotation frequencies for our parameterized EOS, we
used the open-source code \texttt{rns} for axisymmetric rapid rotation in the
updated form \texttt{rns2.0}~\cite{RNS}.  For a given EOS, the model with
maximum spin is ordinarily close to the model with maximum mass, but that need
not be true for EOSs that yield two local mass maxima. The resulting
calculation of maximum rotation requires some care, and the method we use is
described in Appendix B.  The error incurred in using the parameterized EOS
instead of a particular model is 2.7\% (mean+1$\sigma$). 

\begin{figure}[!htb]
\begin{center}
\includegraphics[width=3.4in]{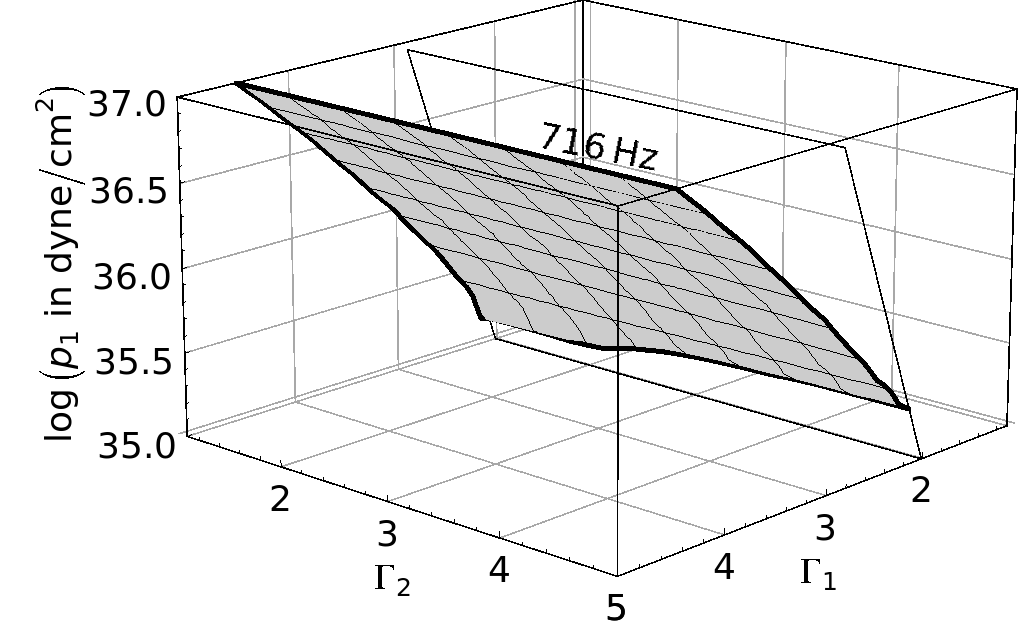}
\caption{The above surface represents the set of parameters that result in a
maximum spin frequency of 716~Hz for the top surface.  For high values of $p_1$
there is no dependence on $\Gamma_3$.
The wedge at the back right is the shaded 
region of Fig.~\ref{fig:candidatefits}, corresponding to 
incompatible values of $p_1$ and $\Gamma_1$.}
\label{fig:f716}
\end{center}
\end{figure}

\begin{figure}[!htb]
\begin{center}
\includegraphics[width=3.4in]{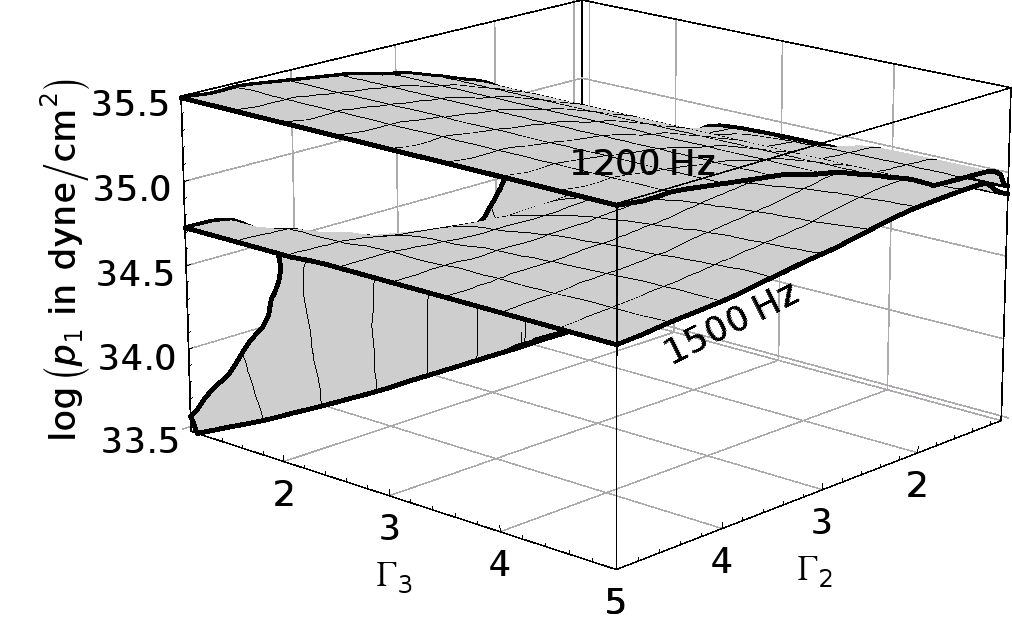}
\caption{The above surfaces represent the set of parameters that result in a
maximum spin frequency of 1200~Hz for the top surface and 1500~Hz for the
bottom surface.  That is, observations of such high spin frequencies would
constrain the EOS to lie below the corresponding surface.
For these surfaces $\Gamma_1=5$, the least constraining value.}
\label{fig:fall}
\end{center}
\end{figure}

Spin frequencies of 716~Hz and even the possible 1122~Hz turn out to be
very weak constraints because both are well below the Kepler frequencies of
most EOSs.  Thus we plot surfaces of parameters giving maximum
rotation frequencies of 716~Hz in Fig.~\ref{fig:f716} and 1300~Hz and 1500~Hz
in Fig.~\ref{fig:fall}.  The region of parameter space above the maximum
observed spin surface is excluded.  In the top figure, maximum mass stars have
central densities below $\rho_2$ so there is no dependence on $\Gamma_3$.  In
the bottom figure the least constraining value of $\Gamma_1 = 5$ is fixed.  The
surface corresponding to a rotation of 716 Hz only constrains the parameter
space that we consider ($p_1<10^{35.5}$~dyne/cm$^3$) if $\Gamma_1 \lesssim 2.5$.
The minimum observed rotation rate necessary to place a firm upper limit on
$p_1$ is roughly 1200~Hz for $\Gamma_1 = 5$.  The surface $f_{\rm max} =
1500$~Hz for $\Gamma_1 = 5$ is also displayed in Fig.~\ref{fig:fall} to
demonstrate that much higher rotation frequencies must be observed in order to
place strong limits on the parameter space.

Because it is computationally expensive to use \texttt{rns} to evaluate the
maximum rotation frequency for a wide range of values in a 4-parameter space,
one can also use an empirical formula.  Haensel and
Zdunik~\cite{HaenselZdunik1989} found that the maximum stable rotation for a
given EOS can be found from the maximum-mass spherically symmetric model for
that EOS with mass $M_{\rm s}$ and radius $R_{\rm s}$ :
\begin{equation}
\left(\frac{\Omega_{\rm max}}{10^4 \, \rm s^{-1}}\right) \approx
\kappa \left(\frac{M_{\rm s}}{M_\odot}\right)^{\frac{1}{2}}
\left(\frac{R_{\rm s}}{10 \, \rm km}\right)^{-\frac{3}{2}}.
\end{equation}
In other words the maximum rotation is proportional to the square root of the
average density of the star.

The original calculation of Haensel and Zdunik gave $\kappa = 0.77$. An
overview of subsequent calculations is given by Haensel et al.
in~\cite{HaenselSalgadoBonazzola1995}, reporting values of $\kappa = 0.76-0.79$
for a range of EOS sets and calculation methods including those of
\cite{FriedmanParkerIpser1986, LattimerPrakashMasakYahil1990,
CookShapiroTeukolsky1994}.  If we calculate maximum rotations with \texttt{rns}
as described above, using the 34 tabulated EOSs, we find
$\kappa = 0.786 \pm 0.030$.  The corresponding best
fit parameterized EOSs give $\kappa = 0.779 \pm 0.027$.

\subsection{Moment of inertia or radius of a neutron star of known mass}
\label{sec:moment}

The moment of inertia of the more massive component, pulsar A, in the double
pulsar PSR~J0737-3039 may be determined to an accuracy of 10\% within the next
few years~\cite{Lattimer:2004nj} by measuring the advance of the system's
periastron, and implications for candidate EOSs have been examined
in~\cite{bejger-bulik-haensel, LattimerSchutz, Morrison2004b}.  As noted
earlier, by finding both mass and moment of inertia of the same star one
imposes a significantly stronger constraint on the EOS parameter space than the
constraints associated with measurements of mass or spin alone: The latter
restrict the EOS to the region of parameter space lying on one side of a
surface, the region associated with the inequality $M_{\rm max}(p_1, \Gamma_i)
> M_{\rm observed}$ or with $\Omega_{\rm max}(p_1, \Gamma_i) > \Omega_{\rm
observed}$.    The simultaneous measurement, on the other hand, restricts the
EOS to a single surface.   That is, in an $n$-dimensional parameter space, the
full n-dimensional set of EOSs which allow a 1.338 $M_\odot$ model, and those
EOSs for which that model has moment of inertia $I_{\rm observed}$ form the
($n-1$)-dimensional surface in parameter space given by $I(p_1, \Gamma_i,
M=1.338M_\odot) = I_{\rm observed}$.  (We use here the fact that the 44 Hz spin
frequency of pulsar A is slow enough that the moment of inertia is nearly that
of the spherical star.) Moreover, for almost all EOSs in the parameter space,
the central density of a  $1.338~M_\odot$ star is below the transition density
$\rho_2$.  Thus the surfaces of constant moment of inertia have negligible
dependence on $\Gamma_3$, the adiabatic index above $\rho_2$, and the EOS is
restricted to the {\it two-dimensional} surface in the
$p_1$-$\Gamma_1$-$\Gamma_2$ space given by $I(p_1, \Gamma_1,\Gamma_2,
M=1.338M_\odot) = I_{\rm observed}$.

This difference in dimensionality means that, in principle, the simultaneous
equalities that give the constraint from observing two features of the same
star are dramatically stronger than the inequalities associated with
measurements of mass or spin alone.  In practice, however, the two-dimensional
constraint surface is thickened by the error of the measurement.   The
additional thickness associated with the error with which the parametrized EOS
can reproduce the moment of inertia of the true EOS is smaller, because the
parameterized EOS reproduces the moment of inertia of the 34 candidate EOSs to
within 2.8\% ($|$mean$|+1\sigma$).  

\begin{figure}[!htb]
\begin{center}
\includegraphics[width=3.4in]{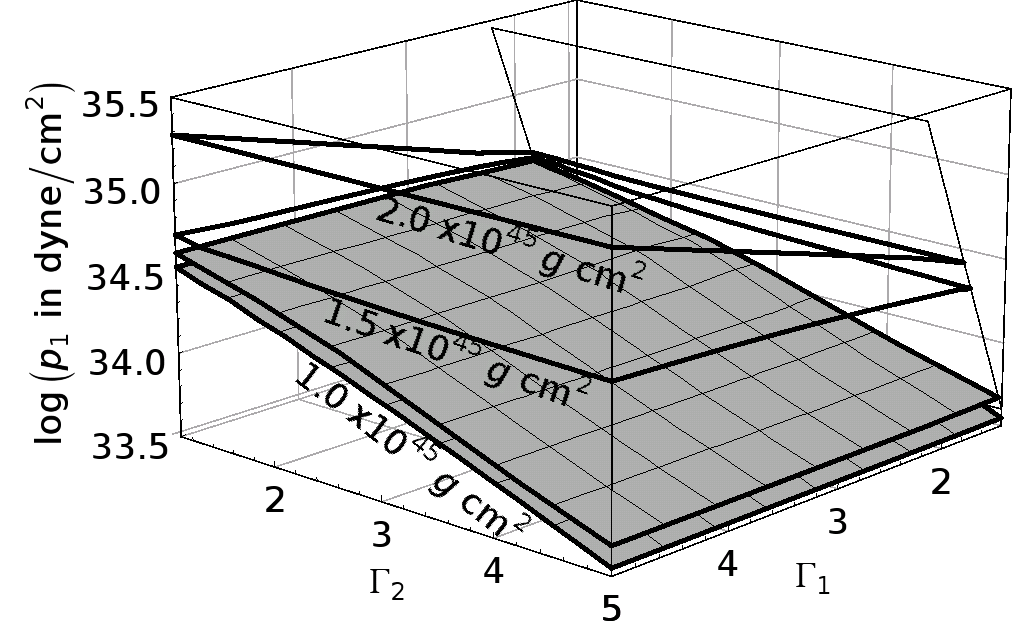}
\caption{The above surfaces represent the set of parameters that result in a
star with a mass of $1.338~M_\odot$ and a fixed moment of inertia, i.e.\
possible near-future measurements of PSR~J0737-3039A.
$I=1.0 \times 10^{45}$~g\,cm$^2$ for the shaded surfaces, whose separation
corresponds to varying $\Gamma_3$.  $I=1.5 \times
10^{45}$~g\,cm$^2$ for the middle outlined surface.  $I=2.0 \times
10^{45}$~g\,cm$^2$ for the top outlined surface.
The wedge at the back right is the shaded 
region of Fig.~\ref{fig:candidatefits}, corresponding to 
incompatible values of $p_1$ and $\Gamma_1$.}
\label{fig:i1338}
\end{center}
\end{figure}

In Fig.~\ref{fig:i1338} we plot surfaces of constant moment of inertia that
span the range associated with the collection of candidate EOSs.  The lower
shaded surface represents $I=1.0 \times 10^{45}$~g\,cm$^2$.  This surface has
very little dependence on $\Gamma_1$ because it represents a more compact star,
and thus for a fixed mass, most of the mass is in a denser state $\rho>\rho_1$.
The structures of these stars do depend on $\Gamma_3$, and the corresponding
dependence of $I$ on $\Gamma_3$ is shown by the 
separation between the surfaces in
Fig.~\ref{fig:i1338}.  The middle outlined surface represents $I=1.5 \times
10^{45}$~g\,cm$^2$, and is almost a surface of constant $p_1$.  The top
outlined surface represents $I=2.0 \times 10^{45}$~g\,cm$^2$.  This surface has
little dependence on $\Gamma_2$, because a star with an EOS on this surface
would be less compact and thus most of its mass would be in a lower density
state $\rho<\rho_1$. 
 
\begin{figure}[!htb]
\begin{center}
\includegraphics[width=3.4in]{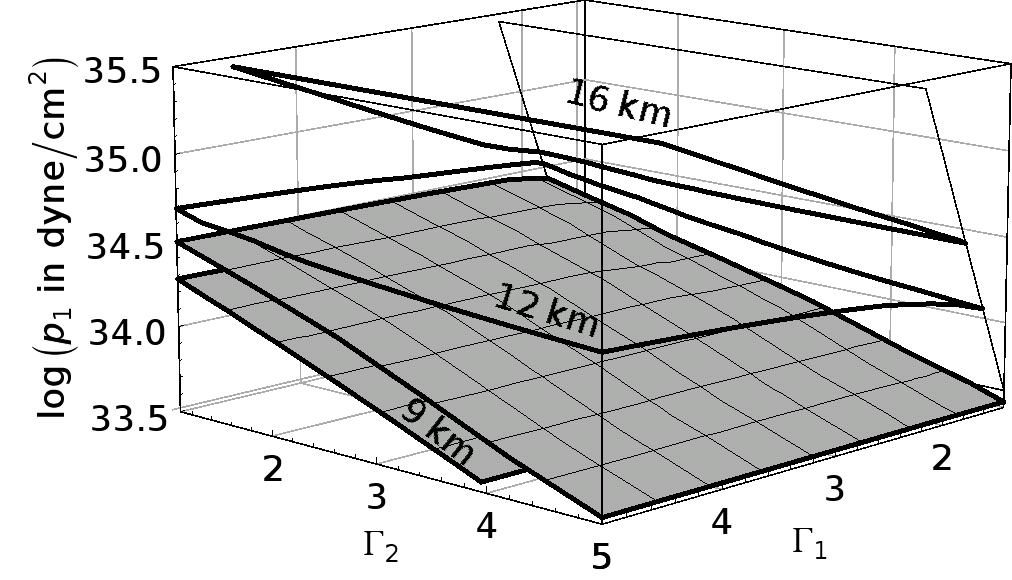}
\caption{The above surfaces represent the set of parameters that result in a
star with a mass of $1.4~M_\odot$ and a fixed radius.
$R=9$~km for the shaded surfaces, whose separation corresponds to varying
$\Gamma_3$.  $R=12$~km for the middle outlined surface.  $R=16$~km for the
top outlined surface.
The wedge at the back right is the shaded 
region of Fig.~\ref{fig:candidatefits}, corresponding to 
incompatible values of $p_1$ and $\Gamma_1$.}
\label{fig:m14r12}
\end{center}
\end{figure}

If the mass of a neutron star is already known, a measurement of the radius
constrains the EOS to a surface of constant mass and radius, $R(p_1,\Gamma_i) =
R_{\rm observed}, M(p_1,\Gamma_i) = M_{\rm observed}$ in the 4-dimensional
parameter space.  The thickness of the surface is dominated by the uncertainty
in the radius and mass measurements, since our parameterization produces the
same radius as the candidate EOSs to within 1.7\% ($|$mean$|+1\sigma$).  We
plot in Fig.~\ref{fig:m14r12} surfaces of constant radius for a $1.4~M_\odot$
star that span the range of radii associated with the collection of candidate
EOSs.  As with the moment of inertia, the radius depends negligibly on
$\Gamma_3$ as long as the radius is greater than 11~km.  For smaller radii, the
variation with $\Gamma_3$ is shown by the separation between the surfaces
in Fig.~\ref{fig:m14r12}. 

Very recently analyses of time-resolved spectroscopic data during
thermonuclear bursts from two neutron stars in low-mass x-ray binaries were
combined with distance estimates to yield $M=1.4~M_\odot$ and $R=11$~km or
$M=1.7~M_\odot$ and $R=9$~km for EXO~1745-248~\cite{Ozel:2008kb} and
$M=1.8~M_\odot$ and $R=10$~km for 4U~1608-52~\cite{Guver:2008gc}, both with
error bars of about 1~km in $R$.
These results are more model dependent than the eventual
measurement of the moment of inertia of PSR~J0737-6069A, 
but the accuracy of the measurement of $I$ remains to be seen.

\subsection{Combining constraints} 
\label{sec:combining}

\begin{figure}[!htb]
\begin{center}
\includegraphics[width=3.4in]{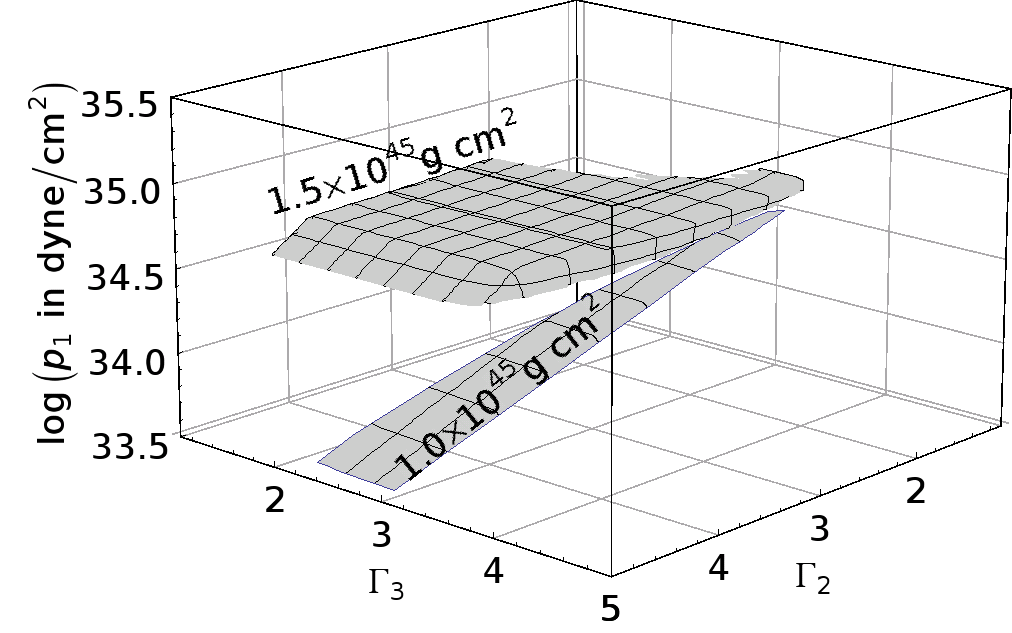}
\includegraphics[width=3.4in]{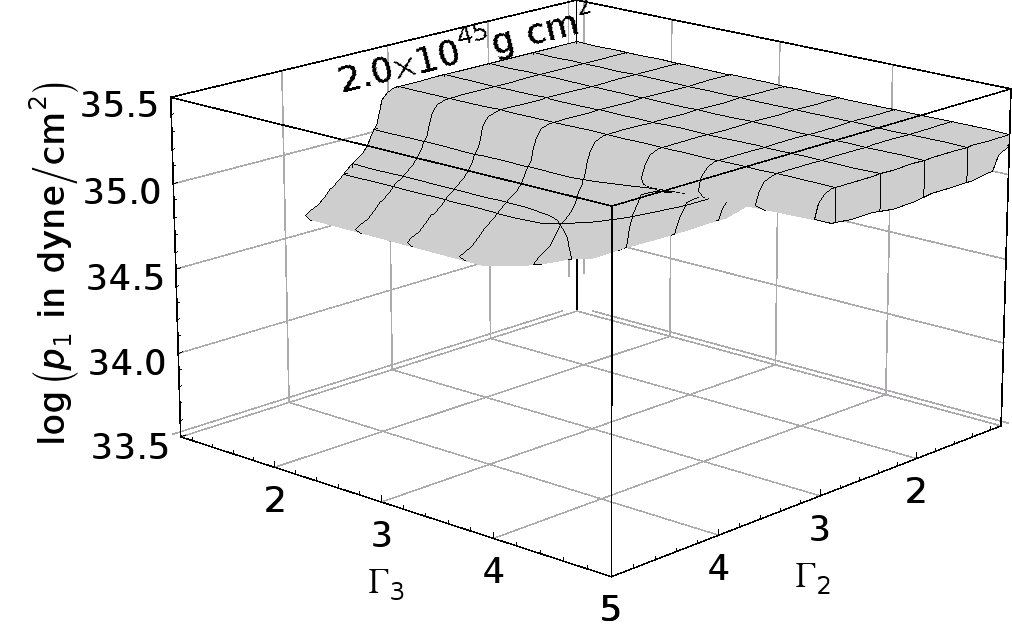}
\caption{The figure portrays the joint constraint imposed by causality ($v_{\rm s,max} < 1+$mean$+1\sigma$), the
existence of a $1.7~M_\odot$ neutron star, and by a future measurement of the
moment of inertia $I$ of J0737-3039A.  Each thick shaded surface is the volume
in $\Gamma_2-\Gamma_3-p_1$ space allowed by the joint constraint for the
labeled value of $I$.} 
\label{fig:imvs3d}
\end{center}
\end{figure}

The simultaneous constraints imposed by causality, a maximum observed mass of
1.7$M_\odot$, and a future measurement of the moment of inertia of
PSR~J0737-3039A, restrict the parameter space to the intersection of the
allowed regions of Figs.~\ref{fig:vsound}, \ref{fig:mall}, and
\ref{fig:i1338}. We show in Fig. \ref{fig:imvs3d} the projection of this
jointly constrained region on the $p_1-\Gamma_2-\Gamma_3$ subspace.  This
allows one to see the cutoffs imposed by causality that eliminate large values
of $\Gamma_2$ and $\Gamma_3$ and (in the top figure) the cutoffs imposed by the
existence of a $1.7M_\odot$ model that eliminates small values of $\Gamma_2$
and $\Gamma_3$.

We noted above that measuring the moment of inertia of a 1.338$M_\odot$ star
restricts the EOS at densities below $\rho_2$ to a two-dimensional surface in
the $p_1-\Gamma_1-\Gamma_2$ space.   In the full 4-dimensional parameter space,
the corresponding surfaces of constant $M$ and $I$ of Fig.~\ref{fig:imvs3d} are
then three dimensional and independent of $\Gamma_3$.  Their projections onto
the $p_1-\Gamma_2-\Gamma_3$ subspace are again three-dimensional and
independent of $\Gamma_3$, their thickness due to the unseen dependence of
the mass and moment of inertia on $\Gamma_1$.  For small moments of inertia
there is negligible dependence on $\Gamma_1$ so the allowed volume in
Fig.~\ref{fig:imvs3d} is thin.  The thickness of the allowed volume increases
as the moment of inertia increases because the dependence on $\Gamma_1$ also
increases.

\begin{figure}[!htb]
\begin{center}
\includegraphics[width=3.4in]{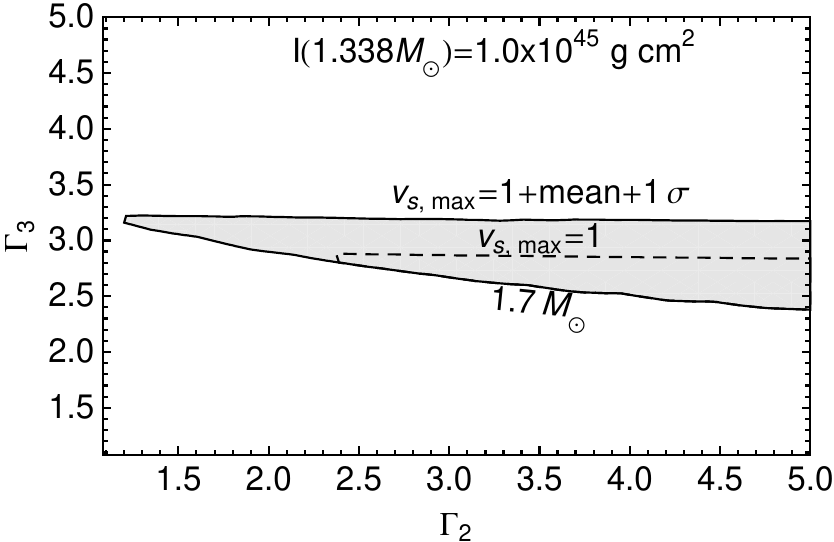}
\includegraphics[width=3.4in]{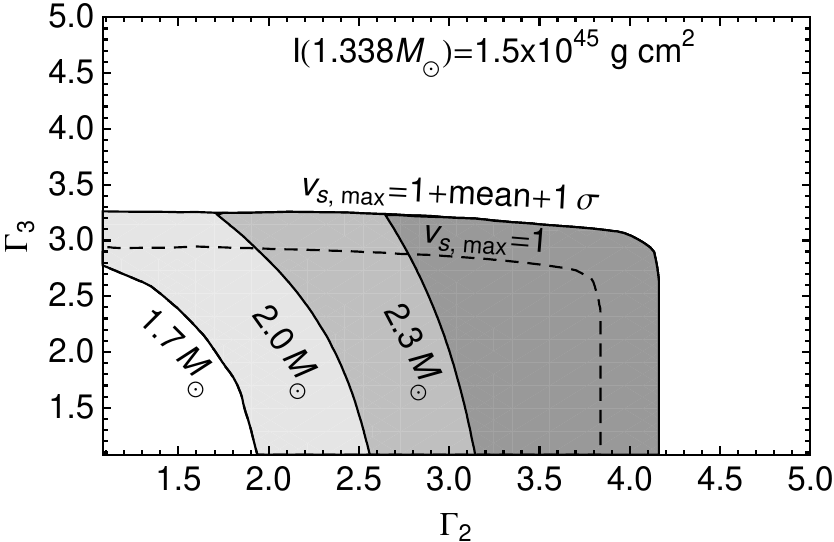}
\includegraphics[width=3.4in]{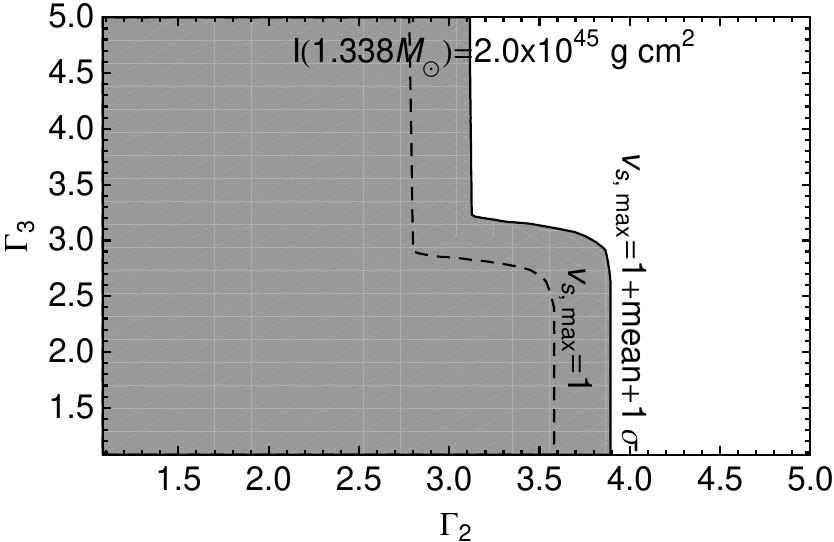}
\caption{The allowed values of $\Gamma_2$ and $\Gamma_3$ depend strongly on the
moment of inertia of PSR~J0737-3039A.  In top, middle and bottom figures,
respectively, $I$ has the values $1.0 \times 10^{45}$~g\,cm$^2$, $I=1.5 \times
10^{45}$~g\,cm$^2$ and $I=2.0 \times 10^{45}$~g\,cm$^2$. In each figure the
upper curves are the $v_{\rm s,max} = 1$ (dotted) and $v_{\rm s,max} = 1+$mean$+1\sigma=1.12$ (solid)
causality constraints. Shading indicates a range of possible maximum mass
constraints, with increasing maximum mass leading to a smaller allowed
area.
All shaded areas are allowed for a $1.7~M_\odot$
 maximum neutron star mass.
The medium and dark shades are allowed if a
$2.0~M_\odot$ star is confirmed.  Only the darkest shade is allowed if a
$2.3~M_\odot$ star is confirmed.
\label{fig:islices}}
\end{center}
\end{figure}

In Fig.~\ref{fig:islices} we explore a relation between the moment of
inertia $I(1.338)$ of PSR~J0737-3039A and the maximum neutron star mass, in
spite of the fact that the maximum mass is significantly greater than
1.338~$M_\odot$.
For three values of the moment of inertia that span the full range
associated with our collection of candidate EOSs, we show joint constraints
on $\Gamma_2$ and $\Gamma_3$ including causality and maximum neutron star
mass.
For $I(1.338)=1.0\times10^{45}$~g\,cm$^2$, $\Gamma_2$ is nearly
unconstrained, while $\Gamma_3$ is required to lie in a small range between
the causality constraint and the reliable observations of stars with mass
1.7~$M_\odot$.
For larger values of $I(1.338)$, $\Gamma_2$ is more constrained and
$\Gamma_3$ is less constrained.
However, the highest values of $I(1.338)$ are associated with the highest
maximum neutron star masses.
Thus, if a neutron star mass of about $2.3~M_\odot$ is confirmed, it
implies that $I(1.338)$ is about $2\times10^{45}$~g\,cm$^2$.
Conversely iff $I(1.338)$ is measured first and is about
$1\times10^{45}$~g\,cm$^2$, it implies that the maximum neutron star mass
is less than about 1.9~$M_\odot$.

\begin{figure}[!htb]
\begin{center}
\includegraphics[width=3.4in]{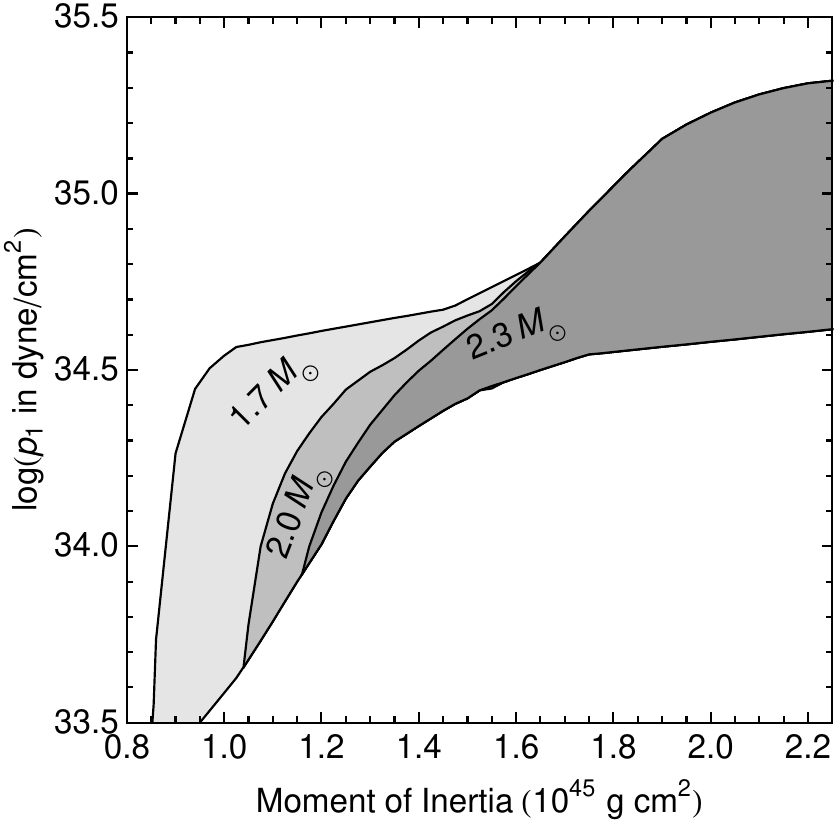}
\caption{The allowed range of $p_1$ as a function of the moment of inertia of
J0737-3039A when combined with causality ($v_{\rm s,max}=1+$mean$+1\sigma$) and
observed mass constraints.  All shaded areas are allowed by a
$1.7~M_\odot$ maximum mass.  The medium and dark shades are allowed if a
$2.0~M_\odot$ star is confirmed.  Only the darkest shade is allowed if a
$2.3~M_\odot$ star is confirmed.} 
\label{fig:p2ofi}
\end{center}
\end{figure}

The allowed range for $p_1$ as a function of the moment of inertia of
J0737-3039A is shown in Fig.~\ref{fig:p2ofi}.  The entire shaded range is
allowed for a $1.7~M_\odot$ maximum mass.
The medium and darker shades are allowed for a $2.0~M_\odot$ maximum mass.  
Only the range with the darker shade is allowed if a
$2.3~M_\odot$ star is confirmed.  It should be noted that for small moments of
inertia, this plot overstates the uncertainty in the allowed parameter range.
As shown in Fig.~\ref{fig:imvs3d}, the allowed volume in
$\Gamma_2-\Gamma_3-p_1$ space for a small moment of inertia observation is
essentially two dimensional.  If the moment of inertia is measured to be this
small, then the EOS would be better parameterized with the linear combination
$\alpha \log(p_1) + \beta \Gamma_2$ instead of two separate parameters
$\log(p_1)$ and $\Gamma_2$. 

\section{Discussion}

We have shown how one can use a parameterized piecewise
polytropic EOS to systematize the study of  observational constraints on the
EOS of cold, high-density matter.  We think that our choice of a 4-parameter
EOS strikes an appropriate balance  between the accuracy of approximation that
a larger number of parameters would  provide and the number of observational
parameters that have been measured or are  likely to be measured in the next
several years.   The simple choice of a piecewise  polytrope, with
discontinuities in the polytropic index, leads to suitable accuracy  in
approximating global features of a star.  But the discontinuity reduces the
expected accuracy with which the parameterized EOS can approximate the local
speed of  sound.  One can largely overcome the problem by using a minor
modification of the parameterized EOS in which a
fixed smoothing function near each dividing density is used to join the two
polytropes.

We see that high-mass neutron stars are likely to provide the
strongest constraints from a single measurement.  The work dramatizes the
significantly more stringent constraints associated with measurements like
this, if two (or more) physical features of the same star  can be measured, and
an $n$-dimensional parameter space is reduced  by one (or more) dimension(s),
to within the error of measurement.  In particular, a moment
of inertia measuremement for PSR~J0737-3039 (whose mass is already precisely
known) could strongly constrain the maximum neutron star mass. 

The effect of EOS-dependent tidal deformation can modify the gravitational
waves produced by inspiraling neutron stars. This modification is largely 
dependent on the radius of the neutron star.   Flanagan and Hinderer
\cite{FlanaganHinderer2007} investigate constraints on an EOS-dependent
tidal parameter, the Love number, from observations of early inspiral. A
companion to this paper~\cite{nrda} uses the parametrized EOS in numerical
simulations to examine the future constraint associated with expected
gravitational-wave observations of late inspiral in binary neutron stars.

Finally, we note that the constraints from observations of different neutron
star populations constrain different density regions of the EOS. For
moderate mass stars such as those found in binary pulsar systems, the EOS
above $\rho_2= 10^{15.0}$~g/cm$^3$ is unimportant. For near-maximum mass
stars, the EOS below $\rho_1=10^{14.7}$~g/cm$^3$ has little effect on
neutron star properties.  This general behavior is independent of the
details of our parameterization.

\acknowledgments 
We thank P. Haensel for helpful suggestions at the start of this work.  J.
Lattimer and M. Alford generously provided EOS tables from
\cite{Lattimer:2000nx} and \cite{AlfordBraby2005hybrid}. Other tables are
from the {\sc LORENE} C++ library ({\tt http://www.lorene.obspm.fr}).  The work
was supported in part by NSF Grants PHY 0503366 and PHY-0555628, by NASA Grant ATP03-0001-0027, and by the Penn State Center for Gravitational-wave Physics under NSF
cooperative agreement PHY-0114375.

\bibliography{eos_tables,recent_eos_ns_lit,polytrope,owen}

\begin{thebibliography}{62}
\expandafter\ifx\csname natexlab\endcsname\relax\def\natexlab#1{#1}\fi
\expandafter\ifx\csname bibnamefont\endcsname\relax
  \def\bibnamefont#1{#1}\fi
\expandafter\ifx\csname bibfnamefont\endcsname\relax
  \def\bibfnamefont#1{#1}\fi
\expandafter\ifx\csname citenamefont\endcsname\relax
  \def\citenamefont#1{#1}\fi
\expandafter\ifx\csname url\endcsname\relax
  \def\url#1{\texttt{#1}}\fi
\expandafter\ifx\csname urlprefix\endcsname\relax\def\urlprefix{URL }\fi
\providecommand{\bibinfo}[2]{#2}
\providecommand{\eprint}[2][]{\url{#2}}

\bibitem[{\citenamefont{{Engvik} et~al.}(1996)\citenamefont{{Engvik}, {Osnes},
  {Hjorth-Jensen}, {Bao}, and {Ostgaard}}}]{EngvikOsnesHjorthJensen1996constr}
\bibinfo{author}{\bibfnamefont{L.}~\bibnamefont{{Engvik}}},
  \bibinfo{author}{\bibfnamefont{E.}~\bibnamefont{{Osnes}}},
  \bibinfo{author}{\bibfnamefont{M.}~\bibnamefont{{Hjorth-Jensen}}},
  \bibinfo{author}{\bibfnamefont{G.}~\bibnamefont{{Bao}}}, \bibnamefont{and}
  \bibinfo{author}{\bibfnamefont{E.}~\bibnamefont{{Ostgaard}}},
  \bibinfo{journal}{\apj} \textbf{\bibinfo{volume}{469}}, \bibinfo{pages}{794}
  (\bibinfo{year}{1996}), \eprint{arXiv:nucl-th/9509016}.

\bibitem[{\citenamefont{{Lattimer} and
  {Prakash}}(2006)}]{lattimerprakash2006nuc}
\bibinfo{author}{\bibfnamefont{J.~M.} \bibnamefont{{Lattimer}}}
  \bibnamefont{and}
  \bibinfo{author}{\bibfnamefont{M.}~\bibnamefont{{Prakash}}},
  \bibinfo{journal}{Nuclear Physics A} \textbf{\bibinfo{volume}{777}},
  \bibinfo{pages}{479} (\bibinfo{year}{2006}).

\bibitem[{\citenamefont{{Kl{\"a}hn} et~al.}(2006)\citenamefont{{Kl{\"a}hn},
  {Blaschke}, {Typel}, {van Dalen}, {Faessler}, {Fuchs}, {Gaitanos},
  {Grigorian}, {Ho}, {Kolomeitsev} et~al.}}]{KlahnBlaschkeTypel2006constraints}
\bibinfo{author}{\bibfnamefont{T.}~\bibnamefont{{Kl{\"a}hn}}},
  \bibinfo{author}{\bibfnamefont{D.}~\bibnamefont{{Blaschke}}},
  \bibinfo{author}{\bibfnamefont{S.}~\bibnamefont{{Typel}}},
  \bibinfo{author}{\bibfnamefont{E.~N.~E.} \bibnamefont{{van Dalen}}},
  \bibinfo{author}{\bibfnamefont{A.}~\bibnamefont{{Faessler}}},
  \bibinfo{author}{\bibfnamefont{C.}~\bibnamefont{{Fuchs}}},
  \bibinfo{author}{\bibfnamefont{T.}~\bibnamefont{{Gaitanos}}},
  \bibinfo{author}{\bibfnamefont{H.}~\bibnamefont{{Grigorian}}},
  \bibinfo{author}{\bibfnamefont{A.}~\bibnamefont{{Ho}}},
  \bibinfo{author}{\bibfnamefont{E.~E.} \bibnamefont{{Kolomeitsev}}},
  \bibnamefont{et~al.}, \bibinfo{journal}{\prc} \textbf{\bibinfo{volume}{74}},
  \bibinfo{pages}{035802} (\bibinfo{year}{2006}),
  \eprint{arXiv:nucl-th/0602038}.

\bibitem[{\citenamefont{{Page} and {Reddy}}(2006)}]{PageReddy2006review}
\bibinfo{author}{\bibfnamefont{D.}~\bibnamefont{{Page}}} \bibnamefont{and}
  \bibinfo{author}{\bibfnamefont{S.}~\bibnamefont{{Reddy}}},
  \bibinfo{journal}{Annual Review of Nuclear and Particle Science}
  \textbf{\bibinfo{volume}{56}}, \bibinfo{pages}{327} (\bibinfo{year}{2006}),
  \eprint{arXiv:astro-ph/0608360}.

\bibitem[{\citenamefont{Lattimer and Prakash}(2001)}]{Lattimer:2000nx}
\bibinfo{author}{\bibfnamefont{J.~M.} \bibnamefont{Lattimer}} \bibnamefont{and}
  \bibinfo{author}{\bibfnamefont{M.}~\bibnamefont{Prakash}},
  \bibinfo{journal}{Astrophys. J.} \textbf{\bibinfo{volume}{550}},
  \bibinfo{pages}{426} (\bibinfo{year}{2001}), \eprint{astro-ph/0002232}.

\bibitem[{\citenamefont{{Vuille} and {Ipser}}(1999)}]{VuilleIpser1999max}
\bibinfo{author}{\bibfnamefont{C.}~\bibnamefont{{Vuille}}} \bibnamefont{and}
  \bibinfo{author}{\bibfnamefont{J.}~\bibnamefont{{Ipser}}}, in
  \emph{\bibinfo{booktitle}{General Relativity and Relativistic Astrophysics}},
  edited by \bibinfo{editor}{\bibfnamefont{C.~P.} \bibnamefont{{Burgess}}}
  \bibnamefont{and} \bibinfo{editor}{\bibfnamefont{R.~C.}
  \bibnamefont{{Myers}}} (\bibinfo{publisher}{American Institute of Physics},
  \bibinfo{address}{College Park MD}, \bibinfo{year}{1999}), vol.
  \bibinfo{volume}{493}, p.~\bibinfo{pages}{60}.

\bibitem[{\citenamefont{{Zdunik} et~al.}(2006)\citenamefont{{Zdunik}, {Bejger},
  {Haensel}, and {Gourgoulhon}}}]{ZdunikBejgerHaensel2006phase}
\bibinfo{author}{\bibfnamefont{J.~L.} \bibnamefont{{Zdunik}}},
  \bibinfo{author}{\bibfnamefont{M.}~\bibnamefont{{Bejger}}},
  \bibinfo{author}{\bibfnamefont{P.}~\bibnamefont{{Haensel}}},
  \bibnamefont{and}
  \bibinfo{author}{\bibfnamefont{E.}~\bibnamefont{{Gourgoulhon}}},
  \bibinfo{journal}{\aap} \textbf{\bibinfo{volume}{450}}, \bibinfo{pages}{747}
  (\bibinfo{year}{2006}), \eprint{arXiv:astro-ph/0509806}.

\bibitem[{\citenamefont{{Bejger}
  et~al.}(2005{\natexlab{a}})\citenamefont{{Bejger}, {Haensel}, and
  {Zdunik}}}]{BejgerHaenselZdunik2005phase}
\bibinfo{author}{\bibfnamefont{M.}~\bibnamefont{{Bejger}}},
  \bibinfo{author}{\bibfnamefont{P.}~\bibnamefont{{Haensel}}},
  \bibnamefont{and} \bibinfo{author}{\bibfnamefont{J.~L.}
  \bibnamefont{{Zdunik}}}, \bibinfo{journal}{\mnras}
  \textbf{\bibinfo{volume}{359}}, \bibinfo{pages}{699}
  (\bibinfo{year}{2005}{\natexlab{a}}), \eprint{arXiv:astro-ph/0502348}.

\bibitem[{\citenamefont{{Haensel} and
  {Potekhin}}(2004)}]{HaenselPotekhin2004anal}
\bibinfo{author}{\bibfnamefont{P.}~\bibnamefont{{Haensel}}} \bibnamefont{and}
  \bibinfo{author}{\bibfnamefont{A.~Y.} \bibnamefont{{Potekhin}}},
  \bibinfo{journal}{\aap} \textbf{\bibinfo{volume}{428}}, \bibinfo{pages}{191}
  (\bibinfo{year}{2004}), \eprint{arXiv:astro-ph/0408324}.

\bibitem[{\citenamefont{Shibata et~al.}(2005)\citenamefont{Shibata, Taniguchi,
  and Uryu}}]{shibata2005es}
\bibinfo{author}{\bibfnamefont{M.}~\bibnamefont{Shibata}},
  \bibinfo{author}{\bibfnamefont{K.}~\bibnamefont{Taniguchi}},
  \bibnamefont{and} \bibinfo{author}{\bibfnamefont{K.}~\bibnamefont{Uryu}},
  \bibinfo{journal}{Physical Review D} \textbf{\bibinfo{volume}{71}},
  \bibinfo{pages}{084021} (\bibinfo{year}{2005}),
  \urlprefix\url{http://www.citebase.org/abstract?id=oai:arXiv.org:gr-qc/05031%
19}.

\bibitem[{\citenamefont{Read et~al.}(2008)\citenamefont{Read, Markakis,
  Shibata, Ury\=u, and Friedman}}]{nrda}
\bibinfo{author}{\bibfnamefont{J.~S.} \bibnamefont{Read}},
  \bibinfo{author}{\bibfnamefont{C.}~\bibnamefont{Markakis}},
  \bibinfo{author}{\bibfnamefont{M.}~\bibnamefont{Shibata}},
  \bibinfo{author}{\bibfnamefont{K.}~\bibnamefont{Ury\=u}}, \bibnamefont{and}
  \bibinfo{author}{\bibfnamefont{J.}~\bibnamefont{Friedman}},
  \bibinfo{journal}{in preparation}  (\bibinfo{year}{2008}).

\bibitem[{\citenamefont{Lackey et~al.}(2006)\citenamefont{Lackey, Nayyar, and
  Owen}}]{Lackey:2005tk}
\bibinfo{author}{\bibfnamefont{B.~D.} \bibnamefont{Lackey}},
  \bibinfo{author}{\bibfnamefont{M.}~\bibnamefont{Nayyar}}, \bibnamefont{and}
  \bibinfo{author}{\bibfnamefont{B.~J.} \bibnamefont{Owen}},
  \bibinfo{journal}{Phys. Rev. D} \textbf{\bibinfo{volume}{73}},
  \bibinfo{pages}{024021} (\bibinfo{year}{2006}), \eprint{astro-ph/0507312}.

\bibitem[{\citenamefont{Lattimer and Schutz}(2005)}]{Lattimer:2004nj}
\bibinfo{author}{\bibfnamefont{J.~M.} \bibnamefont{Lattimer}} \bibnamefont{and}
  \bibinfo{author}{\bibfnamefont{B.~F.} \bibnamefont{Schutz}},
  \bibinfo{journal}{Astrophys. J.} \textbf{\bibinfo{volume}{629}},
  \bibinfo{pages}{979} (\bibinfo{year}{2005}), \eprint{astro-ph/0411470}.

\bibitem[{\citenamefont{{Bejger}
  et~al.}(2005{\natexlab{b}})\citenamefont{{Bejger}, {Bulik}, and
  {Haensel}}}]{bejger-bulik-haensel}
\bibinfo{author}{\bibfnamefont{M.}~\bibnamefont{{Bejger}}},
  \bibinfo{author}{\bibfnamefont{T.}~\bibnamefont{{Bulik}}}, \bibnamefont{and}
  \bibinfo{author}{\bibfnamefont{P.}~\bibnamefont{{Haensel}}},
  \bibinfo{journal}{\mnras} \textbf{\bibinfo{volume}{364}},
  \bibinfo{pages}{635} (\bibinfo{year}{2005}{\natexlab{b}}),
  \eprint{arXiv:astro-ph/0508105}.

\bibitem[{\citenamefont{Prakash et~al.}(1988)\citenamefont{Prakash, Ainsworth,
  and Lattimer}}]{pal}
\bibinfo{author}{\bibfnamefont{M.}~\bibnamefont{Prakash}},
  \bibinfo{author}{\bibfnamefont{T.~L.} \bibnamefont{Ainsworth}},
  \bibnamefont{and} \bibinfo{author}{\bibfnamefont{J.~M.}
  \bibnamefont{Lattimer}}, \bibinfo{journal}{Phys. Rev. Lett.}
  \textbf{\bibinfo{volume}{61}}, \bibinfo{pages}{2518} (\bibinfo{year}{1988}).

\bibitem[{\citenamefont{{Douchin} and {Haensel}}(2001)}]{sly4}
\bibinfo{author}{\bibfnamefont{F.}~\bibnamefont{{Douchin}}} \bibnamefont{and}
  \bibinfo{author}{\bibfnamefont{P.}~\bibnamefont{{Haensel}}},
  \bibinfo{journal}{\aap} \textbf{\bibinfo{volume}{380}}, \bibinfo{pages}{151}
  (\bibinfo{year}{2001}), \eprint{arXiv:astro-ph/0111092}.

\bibitem[{\citenamefont{Akmal et~al.}(1998)\citenamefont{Akmal, Pandharipande,
  and Ravenhall}}]{apr}
\bibinfo{author}{\bibfnamefont{A.}~\bibnamefont{Akmal}},
  \bibinfo{author}{\bibfnamefont{V.~R.} \bibnamefont{Pandharipande}},
  \bibnamefont{and} \bibinfo{author}{\bibfnamefont{D.~G.}
  \bibnamefont{Ravenhall}}, \bibinfo{journal}{Phys. Rev. C}
  \textbf{\bibinfo{volume}{58}}, \bibinfo{pages}{1804} (\bibinfo{year}{1998}).

\bibitem[{\citenamefont{{Friedman} and {Pandharipande}}(1981)}]{fp}
\bibinfo{author}{\bibfnamefont{B.}~\bibnamefont{{Friedman}}} \bibnamefont{and}
  \bibinfo{author}{\bibfnamefont{V.~R.} \bibnamefont{{Pandharipande}}},
  \bibinfo{journal}{Nuclear Physics A} \textbf{\bibinfo{volume}{361}},
  \bibinfo{pages}{502} (\bibinfo{year}{1981}).

\bibitem[{\citenamefont{Wiringa et~al.}(1988)\citenamefont{Wiringa, Fiks, and
  Fabrocini}}]{wff}
\bibinfo{author}{\bibfnamefont{R.~B.} \bibnamefont{Wiringa}},
  \bibinfo{author}{\bibfnamefont{V.}~\bibnamefont{Fiks}}, \bibnamefont{and}
  \bibinfo{author}{\bibfnamefont{A.}~\bibnamefont{Fabrocini}},
  \bibinfo{journal}{Phys. Rev. C} \textbf{\bibinfo{volume}{38}},
  \bibinfo{pages}{1010} (\bibinfo{year}{1988}).

\bibitem[{\citenamefont{{Baldo} et~al.}(1997)\citenamefont{{Baldo}, {Bombaci},
  and {Burgio}}}]{bbb2}
\bibinfo{author}{\bibfnamefont{M.}~\bibnamefont{{Baldo}}},
  \bibinfo{author}{\bibfnamefont{I.}~\bibnamefont{{Bombaci}}},
  \bibnamefont{and} \bibinfo{author}{\bibfnamefont{G.~F.}
  \bibnamefont{{Burgio}}}, \bibinfo{journal}{\aap}
  \textbf{\bibinfo{volume}{328}}, \bibinfo{pages}{274} (\bibinfo{year}{1997}),
  \eprint{arXiv:astro-ph/9707277}.

\bibitem[{\citenamefont{Zuo et~al.}(1999)\citenamefont{Zuo, Bombaci, and
  Lombardo}}]{bpal12}
\bibinfo{author}{\bibfnamefont{W.}~\bibnamefont{Zuo}},
  \bibinfo{author}{\bibfnamefont{I.}~\bibnamefont{Bombaci}}, \bibnamefont{and}
  \bibinfo{author}{\bibfnamefont{U.}~\bibnamefont{Lombardo}},
  \bibinfo{journal}{Phys. Rev. C} \textbf{\bibinfo{volume}{60}},
  \bibinfo{pages}{024605} (\bibinfo{year}{1999}).

\bibitem[{\citenamefont{Engvik et~al.}(1996)\citenamefont{Engvik, Bao,
  Hjorth-Jensen, Osnes, and \O{}stgaard}}]{engvik}
\bibinfo{author}{\bibfnamefont{L.}~\bibnamefont{Engvik}},
  \bibinfo{author}{\bibfnamefont{G.}~\bibnamefont{Bao}},
  \bibinfo{author}{\bibfnamefont{M.}~\bibnamefont{Hjorth-Jensen}},
  \bibinfo{author}{\bibfnamefont{E.}~\bibnamefont{Osnes}}, \bibnamefont{and}
  \bibinfo{author}{\bibfnamefont{E.}~\bibnamefont{\O{}stgaard}},
  \bibinfo{journal}{Astrophysical Journal} \textbf{\bibinfo{volume}{469}},
  \bibinfo{pages}{794} (\bibinfo{year}{1996}),
  \eprint{arXiv:nucl-th/9509016v1},
  \urlprefix\url{http://www.citebase.org/abstract?id=oai:arXiv.org:nucl-th/950%
9016}.

\bibitem[{\citenamefont{{Muther} et~al.}(1987)\citenamefont{{Muther},
  {Prakash}, and {Ainsworth}}}]{mpa1}
\bibinfo{author}{\bibfnamefont{H.}~\bibnamefont{{Muther}}},
  \bibinfo{author}{\bibfnamefont{M.}~\bibnamefont{{Prakash}}},
  \bibnamefont{and} \bibinfo{author}{\bibfnamefont{T.~L.}
  \bibnamefont{{Ainsworth}}}, \bibinfo{journal}{Physics Letters B}
  \textbf{\bibinfo{volume}{199}}, \bibinfo{pages}{469} (\bibinfo{year}{1987}).

\bibitem[{\citenamefont{{M{\"u}ller} and {Serot}}(1996)}]{ms}
\bibinfo{author}{\bibfnamefont{H.}~\bibnamefont{{M{\"u}ller}}}
  \bibnamefont{and} \bibinfo{author}{\bibfnamefont{B.~D.}
  \bibnamefont{{Serot}}}, \bibinfo{journal}{Nuclear Physics A}
  \textbf{\bibinfo{volume}{606}}, \bibinfo{pages}{508} (\bibinfo{year}{1996}),
  \eprint{arXiv:nucl-th/9603037}.

\bibitem[{\citenamefont{{Pandharipande} and {Smith}}(1975)}]{ps}
\bibinfo{author}{\bibfnamefont{V.~R.} \bibnamefont{{Pandharipande}}}
  \bibnamefont{and} \bibinfo{author}{\bibfnamefont{R.~A.}
  \bibnamefont{{Smith}}}, \bibinfo{journal}{Nuclear Physics A}
  \textbf{\bibinfo{volume}{237}}, \bibinfo{pages}{507} (\bibinfo{year}{1975}).

\bibitem[{\citenamefont{Glendenning and Schaffner-Bielich}(1999)}]{schaf}
\bibinfo{author}{\bibfnamefont{N.~K.} \bibnamefont{Glendenning}}
  \bibnamefont{and}
  \bibinfo{author}{\bibfnamefont{J.}~\bibnamefont{Schaffner-Bielich}},
  \bibinfo{journal}{Phys. Rev. C} \textbf{\bibinfo{volume}{60}},
  \bibinfo{pages}{025803} (\bibinfo{year}{1999}).

\bibitem[{\citenamefont{{Balberg} and {Gal}}(1997)}]{balbn1h1}
\bibinfo{author}{\bibfnamefont{S.}~\bibnamefont{{Balberg}}} \bibnamefont{and}
  \bibinfo{author}{\bibfnamefont{A.}~\bibnamefont{{Gal}}},
  \bibinfo{journal}{Nuclear Physics A} \textbf{\bibinfo{volume}{625}},
  \bibinfo{pages}{435} (\bibinfo{year}{1997}), \eprint{arXiv:nucl-th/9704013}.

\bibitem[{\citenamefont{{Glendenning}}(1985)}]{glendnh3}
\bibinfo{author}{\bibfnamefont{N.~K.} \bibnamefont{{Glendenning}}},
  \bibinfo{journal}{\apj} \textbf{\bibinfo{volume}{293}}, \bibinfo{pages}{470}
  (\bibinfo{year}{1985}).

\bibitem[{\citenamefont{Prakash et~al.}(1995)\citenamefont{Prakash, Cooke, and
  Lattimer}}]{pcl}
\bibinfo{author}{\bibfnamefont{M.}~\bibnamefont{Prakash}},
  \bibinfo{author}{\bibfnamefont{J.~R.} \bibnamefont{Cooke}}, \bibnamefont{and}
  \bibinfo{author}{\bibfnamefont{J.~M.} \bibnamefont{Lattimer}},
  \bibinfo{journal}{Phys. Rev. D} \textbf{\bibinfo{volume}{52}},
  \bibinfo{pages}{661} (\bibinfo{year}{1995}).

\bibitem[{\citenamefont{{Alford} et~al.}(2005)\citenamefont{{Alford}, {Braby},
  {Paris}, and {Reddy}}}]{AlfordBraby2005hybrid}
\bibinfo{author}{\bibfnamefont{M.}~\bibnamefont{{Alford}}},
  \bibinfo{author}{\bibfnamefont{M.}~\bibnamefont{{Braby}}},
  \bibinfo{author}{\bibfnamefont{M.}~\bibnamefont{{Paris}}}, \bibnamefont{and}
  \bibinfo{author}{\bibfnamefont{S.}~\bibnamefont{{Reddy}}},
  \bibinfo{journal}{\apj} \textbf{\bibinfo{volume}{629}}, \bibinfo{pages}{969}
  (\bibinfo{year}{2005}), \eprint{arXiv:nucl-th/0411016}.

\bibitem[{\citenamefont{Pippard}(1964)}]{pippard}
\bibinfo{author}{\bibfnamefont{A.}~\bibnamefont{Pippard}}
  (\bibinfo{publisher}{Cambridge University Press}, \bibinfo{year}{1964}).

\bibitem[{\citenamefont{Negele and Vautherin}(1973)}]{negele-vautherin}
\bibinfo{author}{\bibfnamefont{J.}~\bibnamefont{Negele}} \bibnamefont{and}
  \bibinfo{author}{\bibfnamefont{D.}~\bibnamefont{Vautherin}},
  \bibinfo{journal}{Nuclear Physics A} \textbf{\bibinfo{volume}{207}},
  \bibinfo{pages}{298} (\bibinfo{year}{1973}).

\bibitem[{\citenamefont{Glendenning and Kettner}(2000)}]{Glendenning:1998ag}
\bibinfo{author}{\bibfnamefont{N.~K.} \bibnamefont{Glendenning}}
  \bibnamefont{and} \bibinfo{author}{\bibfnamefont{C.}~\bibnamefont{Kettner}},
  \bibinfo{journal}{Astron. Astrophys.} \textbf{\bibinfo{volume}{353}},
  \bibinfo{pages}{L9} (\bibinfo{year}{2000}), \eprint{astro-ph/9807155}.

\bibitem[{\citenamefont{Lattimer and Prakash}(2007)}]{Lattimer:2006xb}
\bibinfo{author}{\bibfnamefont{J.~M.} \bibnamefont{Lattimer}} \bibnamefont{and}
  \bibinfo{author}{\bibfnamefont{M.}~\bibnamefont{Prakash}},
  \bibinfo{journal}{Phys. Rept.} \textbf{\bibinfo{volume}{442}},
  \bibinfo{pages}{109} (\bibinfo{year}{2007}), \eprint{astro-ph/0612440}.

\bibitem[{\citenamefont{Ransom et~al.}(2005)}]{Ransom:2005ae}
\bibinfo{author}{\bibfnamefont{S.~M.} \bibnamefont{Ransom}}
  \bibnamefont{et~al.}, \bibinfo{journal}{Science}
  \textbf{\bibinfo{volume}{307}}, \bibinfo{pages}{892} (\bibinfo{year}{2005}),
  \eprint{astro-ph/0501230}.

\bibitem[{\citenamefont{{Freire} et~al.}(2008)\citenamefont{{Freire},
  {Wolszczan}, {van den Berg}, and {Hessels}}}]{Freire:2007xg}
\bibinfo{author}{\bibfnamefont{P.~C.~C.} \bibnamefont{{Freire}}},
  \bibinfo{author}{\bibfnamefont{A.}~\bibnamefont{{Wolszczan}}},
  \bibinfo{author}{\bibfnamefont{M.}~\bibnamefont{{van den Berg}}},
  \bibnamefont{and} \bibinfo{author}{\bibfnamefont{J.~W.~T.}
  \bibnamefont{{Hessels}}}, \bibinfo{journal}{\apj}
  \textbf{\bibinfo{volume}{679}}, \bibinfo{pages}{1433} (\bibinfo{year}{2008}),
  \eprint{0712.3826}.

\bibitem[{\citenamefont{{Champion} et~al.}(2008)\citenamefont{{Champion},
  {Ransom}, {Lazarus}, {Camilo}, {Bassa}, {Kaspi}, {Nice}, {Freire}, {Stairs},
  {van Leeuwen} et~al.}}]{Champion:2008ge}
\bibinfo{author}{\bibfnamefont{D.~J.} \bibnamefont{{Champion}}},
  \bibinfo{author}{\bibfnamefont{S.~M.} \bibnamefont{{Ransom}}},
  \bibinfo{author}{\bibfnamefont{P.}~\bibnamefont{{Lazarus}}},
  \bibinfo{author}{\bibfnamefont{F.}~\bibnamefont{{Camilo}}},
  \bibinfo{author}{\bibfnamefont{C.}~\bibnamefont{{Bassa}}},
  \bibinfo{author}{\bibfnamefont{V.~M.} \bibnamefont{{Kaspi}}},
  \bibinfo{author}{\bibfnamefont{D.~J.} \bibnamefont{{Nice}}},
  \bibinfo{author}{\bibfnamefont{P.~C.~C.} \bibnamefont{{Freire}}},
  \bibinfo{author}{\bibfnamefont{I.~H.} \bibnamefont{{Stairs}}},
  \bibinfo{author}{\bibfnamefont{J.}~\bibnamefont{{van Leeuwen}}},
  \bibnamefont{et~al.}, \bibinfo{journal}{Science}
  \textbf{\bibinfo{volume}{320}}, \bibinfo{pages}{1309} (\bibinfo{year}{2008}),
  \eprint{0805.2396}.

\bibitem[{\citenamefont{{Verbiest} et~al.}(2008)\citenamefont{{Verbiest},
  {Bailes}, {van Straten}, {Hobbs}, {Edwards}, {Manchester}, {Bhat},
  {Sarkissian}, {Jacoby}, and {Kulkarni}}}]{Verbiest:2008gy}
\bibinfo{author}{\bibfnamefont{J.~P.~W.} \bibnamefont{{Verbiest}}},
  \bibinfo{author}{\bibfnamefont{M.}~\bibnamefont{{Bailes}}},
  \bibinfo{author}{\bibfnamefont{W.}~\bibnamefont{{van Straten}}},
  \bibinfo{author}{\bibfnamefont{G.~B.} \bibnamefont{{Hobbs}}},
  \bibinfo{author}{\bibfnamefont{R.~T.} \bibnamefont{{Edwards}}},
  \bibinfo{author}{\bibfnamefont{R.~N.} \bibnamefont{{Manchester}}},
  \bibinfo{author}{\bibfnamefont{N.~D.~R.} \bibnamefont{{Bhat}}},
  \bibinfo{author}{\bibfnamefont{J.~M.} \bibnamefont{{Sarkissian}}},
  \bibinfo{author}{\bibfnamefont{B.~A.} \bibnamefont{{Jacoby}}},
  \bibnamefont{and} \bibinfo{author}{\bibfnamefont{S.~R.}
  \bibnamefont{{Kulkarni}}}, \bibinfo{journal}{\apj}
  \textbf{\bibinfo{volume}{679}}, \bibinfo{pages}{675} (\bibinfo{year}{2008}),
  \eprint{0801.2589}.

\bibitem[{\citenamefont{Freire et~al.}(2008)}]{Freire:2007jd}
\bibinfo{author}{\bibfnamefont{P.~C.~C.} \bibnamefont{Freire}}
  \bibnamefont{et~al.}, \bibinfo{journal}{Astrophys. J.}
  \textbf{\bibinfo{volume}{675}}, \bibinfo{pages}{670} (\bibinfo{year}{2008}),
  \eprint{arXiv:0711.0925 [astro-ph]}.

\bibitem[{\citenamefont{Haensel et~al.}(2002)\citenamefont{Haensel, Zdunik, and
  Douchin}}]{hzd02}
\bibinfo{author}{\bibfnamefont{P.}~\bibnamefont{Haensel}},
  \bibinfo{author}{\bibfnamefont{J.}~\bibnamefont{Zdunik}}, \bibnamefont{and}
  \bibinfo{author}{\bibfnamefont{F.}~\bibnamefont{Douchin}},
  \bibinfo{journal}{Astronomy and Astrophysics} \textbf{\bibinfo{volume}{385}},
  \bibinfo{pages}{301} (\bibinfo{year}{2002}), \eprint{astro-ph/0201434}.

\bibitem[{\citenamefont{Cottam et~al.}(2002)\citenamefont{Cottam, Paerels, and
  Mendez}}]{Cottam:2002cu}
\bibinfo{author}{\bibfnamefont{J.}~\bibnamefont{Cottam}},
  \bibinfo{author}{\bibfnamefont{F.}~\bibnamefont{Paerels}}, \bibnamefont{and}
  \bibinfo{author}{\bibfnamefont{M.}~\bibnamefont{Mendez}},
  \bibinfo{journal}{Nature} \textbf{\bibinfo{volume}{420}}, \bibinfo{pages}{51}
  (\bibinfo{year}{2002}), \eprint{astro-ph/0211126}.

\bibitem[{\citenamefont{Cottam et~al.}(2008)}]{Cottam:2007cd}
\bibinfo{author}{\bibfnamefont{J.}~\bibnamefont{Cottam}} \bibnamefont{et~al.},
  \bibinfo{journal}{Astrophys. J.} \textbf{\bibinfo{volume}{672}},
  \bibinfo{pages}{504} (\bibinfo{year}{2008}), \eprint{arXiv:0709.4062
  [astro-ph]}.

\bibitem[{\citenamefont{Ozel}(2006)}]{Ozel:2006bv}
\bibinfo{author}{\bibfnamefont{F.}~\bibnamefont{Ozel}},
  \bibinfo{journal}{Nature} \textbf{\bibinfo{volume}{441}},
  \bibinfo{pages}{1115} (\bibinfo{year}{2006}).

\bibitem[{\citenamefont{Hessels et~al.}(2006)}]{Hessels:2006ze}
\bibinfo{author}{\bibfnamefont{J.~W.~T.} \bibnamefont{Hessels}}
  \bibnamefont{et~al.}, \bibinfo{journal}{Science}
  \textbf{\bibinfo{volume}{311}}, \bibinfo{pages}{1901} (\bibinfo{year}{2006}),
  \eprint{astro-ph/0601337}.

\bibitem[{\citenamefont{Kaaret et~al.}(2007)}]{Kaaret:2006gr}
\bibinfo{author}{\bibfnamefont{P.}~\bibnamefont{Kaaret}} \bibnamefont{et~al.},
  \bibinfo{journal}{Astrophys. J.} \textbf{\bibinfo{volume}{657}},
  \bibinfo{pages}{L97} (\bibinfo{year}{2007}), \eprint{astro-ph/0611716}.

\bibitem[{\citenamefont{Stergioulas et~al.}(2002)\citenamefont{Stergioulas,
  Koranda, and Friedman}}]{skf97}
\bibinfo{author}{\bibfnamefont{N.}~\bibnamefont{Stergioulas}},
  \bibinfo{author}{\bibfnamefont{S.}~\bibnamefont{Koranda}}, \bibnamefont{and}
  \bibinfo{author}{\bibfnamefont{J.}~\bibnamefont{Friedman}},
  \bibinfo{journal}{Ap. J.} \textbf{\bibinfo{volume}{488}},
  \bibinfo{pages}{301} (\bibinfo{year}{2002}).

\bibitem[{\citenamefont{Glendenning}(1992)}]{glendenning92}
\bibinfo{author}{\bibfnamefont{N.}~\bibnamefont{Glendenning}},
  \bibinfo{journal}{Phys. Rev. D} \textbf{\bibinfo{volume}{46}},
  \bibinfo{pages}{4161} (\bibinfo{year}{1992}).

\bibitem[{\citenamefont{Stergioulas}(2000)}]{RNS}
\bibinfo{author}{\bibfnamefont{N.}~\bibnamefont{Stergioulas}},
  \emph{\bibinfo{title}{http://www.gravity.phys.uwm.edu/rns}}
  (\bibinfo{publisher}{UWM Centre for Gravitation and Cosmology},
  \bibinfo{year}{2000}).

\bibitem[{\citenamefont{{Haensel} and {Zdunik}}(1989)}]{HaenselZdunik1989}
\bibinfo{author}{\bibfnamefont{P.}~\bibnamefont{{Haensel}}} \bibnamefont{and}
  \bibinfo{author}{\bibfnamefont{J.~L.} \bibnamefont{{Zdunik}}},
  \bibinfo{journal}{\nat} \textbf{\bibinfo{volume}{340}}, \bibinfo{pages}{617}
  (\bibinfo{year}{1989}).

\bibitem[{\citenamefont{{Haensel} et~al.}(1995)\citenamefont{{Haensel},
  {Salgado}, and {Bonazzola}}}]{HaenselSalgadoBonazzola1995}
\bibinfo{author}{\bibfnamefont{P.}~\bibnamefont{{Haensel}}},
  \bibinfo{author}{\bibfnamefont{M.}~\bibnamefont{{Salgado}}},
  \bibnamefont{and}
  \bibinfo{author}{\bibfnamefont{S.}~\bibnamefont{{Bonazzola}}},
  \bibinfo{journal}{\aap} \textbf{\bibinfo{volume}{296}}, \bibinfo{pages}{745}
  (\bibinfo{year}{1995}).

\bibitem[{\citenamefont{{Friedman} et~al.}(1986)\citenamefont{{Friedman},
  {Parker}, and {Ipser}}}]{FriedmanParkerIpser1986}
\bibinfo{author}{\bibfnamefont{J.~L.} \bibnamefont{{Friedman}}},
  \bibinfo{author}{\bibfnamefont{L.}~\bibnamefont{{Parker}}}, \bibnamefont{and}
  \bibinfo{author}{\bibfnamefont{J.~R.} \bibnamefont{{Ipser}}},
  \bibinfo{journal}{\apj} \textbf{\bibinfo{volume}{304}}, \bibinfo{pages}{115}
  (\bibinfo{year}{1986}).

\bibitem[{\citenamefont{{Lattimer} et~al.}(1990)\citenamefont{{Lattimer},
  {Prakash}, {Masak}, and {Yahil}}}]{LattimerPrakashMasakYahil1990}
\bibinfo{author}{\bibfnamefont{J.~M.} \bibnamefont{{Lattimer}}},
  \bibinfo{author}{\bibfnamefont{M.}~\bibnamefont{{Prakash}}},
  \bibinfo{author}{\bibfnamefont{D.}~\bibnamefont{{Masak}}}, \bibnamefont{and}
  \bibinfo{author}{\bibfnamefont{A.}~\bibnamefont{{Yahil}}},
  \bibinfo{journal}{\apj} \textbf{\bibinfo{volume}{355}}, \bibinfo{pages}{241}
  (\bibinfo{year}{1990}).

\bibitem[{\citenamefont{{Cook} et~al.}(1994)\citenamefont{{Cook}, {Shapiro},
  and {Teukolsky}}}]{CookShapiroTeukolsky1994}
\bibinfo{author}{\bibfnamefont{G.~B.} \bibnamefont{{Cook}}},
  \bibinfo{author}{\bibfnamefont{S.~L.} \bibnamefont{{Shapiro}}},
  \bibnamefont{and} \bibinfo{author}{\bibfnamefont{S.~A.}
  \bibnamefont{{Teukolsky}}}, \bibinfo{journal}{\apj}
  \textbf{\bibinfo{volume}{424}}, \bibinfo{pages}{823} (\bibinfo{year}{1994}).

\bibitem[{\citenamefont{{Lattimer} and {Schutz}}(2005)}]{LattimerSchutz}
\bibinfo{author}{\bibfnamefont{J.~M.} \bibnamefont{{Lattimer}}}
  \bibnamefont{and} \bibinfo{author}{\bibfnamefont{B.~F.}
  \bibnamefont{{Schutz}}}, \bibinfo{journal}{\apj}
  \textbf{\bibinfo{volume}{629}}, \bibinfo{pages}{979} (\bibinfo{year}{2005}),
  \eprint{arXiv:astro-ph/0411470}.

\bibitem[{\citenamefont{{Morrison} et~al.}(2004)\citenamefont{{Morrison},
  {Baumgarte}, {Shapiro}, and {Pandharipande}}}]{Morrison2004b}
\bibinfo{author}{\bibfnamefont{I.~A.} \bibnamefont{{Morrison}}},
  \bibinfo{author}{\bibfnamefont{T.~W.} \bibnamefont{{Baumgarte}}},
  \bibinfo{author}{\bibfnamefont{S.~L.} \bibnamefont{{Shapiro}}},
  \bibnamefont{and} \bibinfo{author}{\bibfnamefont{V.~R.}
  \bibnamefont{{Pandharipande}}}, \bibinfo{journal}{\apjl}
  \textbf{\bibinfo{volume}{617}}, \bibinfo{pages}{L135} (\bibinfo{year}{2004}),
  \eprint{arXiv:astro-ph/0411353}.

\bibitem[{\citenamefont{Ozel et~al.}(2008)\citenamefont{Ozel, Guver, and
  Psaltis}}]{Ozel:2008kb}
\bibinfo{author}{\bibfnamefont{F.}~\bibnamefont{Ozel}},
  \bibinfo{author}{\bibfnamefont{T.}~\bibnamefont{Guver}}, \bibnamefont{and}
  \bibinfo{author}{\bibfnamefont{D.}~\bibnamefont{Psaltis}}
  (\bibinfo{year}{2008}), \eprint{0810.1521}.

\bibitem[{\citenamefont{Guver et~al.}(2008)\citenamefont{Guver, Ozel,
  Cabrera-Lavers, and Wroblewski}}]{Guver:2008gc}
\bibinfo{author}{\bibfnamefont{T.}~\bibnamefont{Guver}},
  \bibinfo{author}{\bibfnamefont{F.}~\bibnamefont{Ozel}},
  \bibinfo{author}{\bibfnamefont{A.}~\bibnamefont{Cabrera-Lavers}},
  \bibnamefont{and}
  \bibinfo{author}{\bibfnamefont{P.}~\bibnamefont{Wroblewski}}
  (\bibinfo{year}{2008}), \eprint{0811.3979}.

\bibitem[{\citenamefont{{Flanagan} and
  {Hinderer}}(2008)}]{FlanaganHinderer2007}
\bibinfo{author}{\bibfnamefont{{\'E}.~{\'E}.} \bibnamefont{{Flanagan}}}
  \bibnamefont{and}
  \bibinfo{author}{\bibfnamefont{T.}~\bibnamefont{{Hinderer}}},
  \bibinfo{journal}{\prd} \textbf{\bibinfo{volume}{77}},
  \bibinfo{pages}{021502} (\bibinfo{year}{2008}), \eprint{0709.1915}.

\bibitem[{\citenamefont{Hartle}(1967)}]{Hartle:1967he}
\bibinfo{author}{\bibfnamefont{J.~B.} \bibnamefont{Hartle}},
  \bibinfo{journal}{Astrophys. J.} \textbf{\bibinfo{volume}{150}},
  \bibinfo{pages}{1005} (\bibinfo{year}{1967}).

\bibitem[{\citenamefont{Lindblom}(1992)}]{lindblom1992es}
\bibinfo{author}{\bibfnamefont{L.}~\bibnamefont{Lindblom}},
  \bibinfo{journal}{The Astrophysical Journal} \textbf{\bibinfo{volume}{308}},
  \bibinfo{pages}{569} (\bibinfo{year}{1992}).

\bibitem[{\citenamefont{{Friedman} et~al.}(1988)\citenamefont{{Friedman},
  {Ipser}, and {Sorkin}}}]{FriedmanIpserSorkin1988}
\bibinfo{author}{\bibfnamefont{J.~L.} \bibnamefont{{Friedman}}},
  \bibinfo{author}{\bibfnamefont{J.~R.} \bibnamefont{{Ipser}}},
  \bibnamefont{and} \bibinfo{author}{\bibfnamefont{R.~D.}
  \bibnamefont{{Sorkin}}}, \bibinfo{journal}{\apj}
  \textbf{\bibinfo{volume}{325}}, \bibinfo{pages}{722} (\bibinfo{year}{1988}).

\bibitem[{\citenamefont{Cook et~al.}(1992)\citenamefont{Cook, Shapiro, and
  Teukolsky}}]{CookShapiroTeukolsky}
\bibinfo{author}{\bibfnamefont{G.}~\bibnamefont{Cook}},
  \bibinfo{author}{\bibfnamefont{S.}~\bibnamefont{Shapiro}}, \bibnamefont{and}
  \bibinfo{author}{\bibfnamefont{S.}~\bibnamefont{Teukolsky}},
  \bibinfo{journal}{Ap. J.} \textbf{\bibinfo{volume}{424}},
  \bibinfo{pages}{823} (\bibinfo{year}{1992}).

\end{thebibliography}

\appendix
\section{Evaluating mass, radius, and moment of inertia}

The moment of inertia of a rotating star is the ratio $I=J/\Omega$, with $J$
the asymptotically defined angular momentum.    In finding the moment of
inertia of spherical models, we use Hartle's slow-rotation
equations~\cite{Hartle:1967he}, adapted
to piecewise polytropes in a way we describe below.  The metric of a slowly
rotating star has to order $\Omega$ the form 
\begin{eqnarray}
ds^2 &=& -e^{2 \Phi(r)} dt^2+e^{2 \lambda(r)} dr^2 -2 \omega(r) r^2 \sin^2 \theta d\phi dt \nonumber \\
     & & +r^2 d\theta^2+r^2 \sin^2 \theta d\phi^2,
\end{eqnarray}
where $\Phi$ and $\lambda$ are the metric functions of the spherical star,
given by \begin{eqnarray}
 e^{2 \lambda(r)} &=& \left(1-\frac{2m(r)}{r}\right)^{-1},\\
 \frac{d\Phi}{dr} &=& - \frac{1}{\epsilon+p}\frac{dp}{dr}, 
 \label{eq:dphidr}\\
 \frac{dp}{dr} &=& -(\epsilon+p)\frac{m+4\pi r^3 p}{r(r-2m)}, 
 \label{eq:dpdr}\\
\frac{dm}{dr} &=& 4 \pi r^2 \epsilon.
\label{eq:dmdr}
\end{eqnarray}
The frame-dragging $\omega(r)$ is obtained from the $t\phi$ component of the
Einstein equation in the form 
\begin{equation}
\frac{1}{r^4}\frac{d}{dr}\left(r^4j\frac{d\overline{\omega}}{dr}\right)+\frac{4}{r}\frac{dj}{dr}\overline{\omega} = 0,
\label{eq:dodr}
\end{equation}
where $\overline{\omega}=\Omega-\omega$ is the angular velocity of the star
measured by a zero-angular-momentum observer and 
\begin{equation}
j(r)=e^{-\Phi}\left(1-\frac{2m}{r}\right)^{1/2}.
\end{equation}
The angular momentum is obtained from $\omega$, which has outside the star the
form $\omega=2J/r^3$.   

In adapting these equations, we roughly follow Lindblom~\cite{lindblom1992es},
replacing $r$ as a radial variable by a generalization $\eta := h-1$ of the
Newtonian enthalpy. \footnote{Lindblom, however, uses $\log h$ instead of
$h-1$ as his radial variable.  Because of the form of the piecewise polytrope,
$h-1$ is a more convenient choice here. This variable is also used by Haensel
and Potekhin in \cite{HaenselPotekhin2004anal}.} Because $\eta$ is monotonic in
$r$, one can integrate outward from its central value to the surface, where
$\eta=0$.  

For the piecewise polytropes of Sec.~\ref{sec:ppdef}, the equation of state
given in terms of $\eta$ is
\begin{eqnarray}
\rho(\eta) & = & \left( \frac{\eta - a_i}{K_i (n_i+1)} \right)^{n_i}, \\
p(\eta) & = & K_i \left( \frac{\eta - a_i}{K_i (n_i+1)} \right)^{n_i+1}, \\
\epsilon(\eta) & = & \rho(\eta) \left( 1 + \frac{a_i + n_i \eta}{n_i + 1}
\right), 
\end{eqnarray}
where $n_i = 1/(\Gamma_i - 1)$ is the polytropic index.

This replacement exploits the first integral $he^\Phi=\sqrt{1-2M/R}$ of the
equation of hydrostatic equilibrium to eliminate the differential equation
~(\ref{eq:dphidr}) for $\Phi$; and the enthalpy, unlike $\epsilon$ and $p$, is
smooth at the surface for a polytropic EOS.
Eqs.~(\ref{eq:dpdr}-\ref{eq:dodr}) are then equivalent to the first-order set
\begin{eqnarray}
\frac{dr}{d\eta} &=& - \frac{r (r - 2m)}{m + 4 \pi r^3 p(\eta)} \frac{1}{\eta + 1} \\
\frac{dm}{d\eta} &=& 4 \pi r^2 \epsilon(\eta) \frac{dr}{d\eta} \\
\frac{d\overline{\omega}}{d\eta} &=& \alpha \frac{dr}{d\eta} \\
\frac{d\alpha}{d\eta} &=& \left[-\frac{4\alpha}{r}+\frac{4\pi(\epsilon+p)(r\alpha+4\overline{\omega})}{1-2m/r} \right]\frac{dr}{d\eta}
\end{eqnarray}
where $\alpha:=d\overline{\omega}/dr$.  

The integration to find the mass, radius, and moment of inertia for a star with
given central value $\eta=\eta_c$ proceeds as follows: Use the initial
conditions $r(\eta_c)=m(\eta_c)=\alpha(\eta_c)=0$ and {\em arbitrarily} choose
a central value $\overline{\omega}_0$ of $\overline{\omega}$.  Integrate to the
surface where $\eta=0$, to obtain the radius $R=r(\eta=0)$ and mass
$M=m(\eta=0)$.  The angular momentum $J$ is found from the radial derivative of
the equation 
\begin{equation}
\overline{\omega} = \Omega - \frac{2 J}{r^3}, 
\end{equation}
evaluated at $r=R$, namely
\be
J = \frac{1}{6}R^4 \alpha(R),
\ee 
and $\Omega$ is then given by
\be
\Omega = \overline{\omega}(R)+\frac{2J}{R^3}.
\ee
These values of $\Omega$ and $J$ are each proportional to the arbitrarily
chosen $\overline\omega_0$, implying that the moment of inertia $J/\Omega$ is
independent of $\overline\omega_0$.

\section{Stability of rotating models}
The mass-shed limit gives a maximally rotating equilibrium model for each
central energy density $\epsilon_c$, but, as in the spherical case, these
equilibrium models are not guaranteed to be stable to perturbations. 

%The equilibrium rotating models for a given cold EOS form a two-parameter
%family, which can be specified by the central energy density $\epsilon_c$ and
%the axis ratio $\mathfrak{r}$, as in the base \texttt{rns} routine, or
%equivalently by $\epsilon_c$ and the angular velocity $\Omega$. In the limit
%that $\Omega \to 0$ and thus $\mathfrak{r} \to 1$, we recover the
%one-parameter family of nonrotating spherical models.

Overall stability in uniformly rotating models is governed by the stability of
the model to pseudoradial perturbations.  As in the spherical $\Omega=0$ case,
there can exist alternating regions of stable and unstable rotating models
along a sequence of fixed $\Omega$. A criterion for the onset of instability is
developed by Friedman, Ipser and Sorkin in \cite{FriedmanIpserSorkin1988}: The
critical points that potentially indicate a change in stability are extrema of
mass-energy $M$
%, where \begin{equation} dM = \mu dM_b + \Omega dJ \end{equation} is a local
%extremum 
under variation in both baryon mass $M_b$ and angular momentum $J$; and can be
determined by extremizing $M_b$ on sequences of constant $J$ or extremising $J$
on sequences of constant $M_b$.  Universally valid searches for limiting
stability, as in for example \cite{ CookShapiroTeukolsky}, have therefore
required explicitly covering the set of models with sequences of constant rest
mass $M_b$ and extremizing $J$ on each one, or vice versa---a computationally
expensive procedure.

\begin{figure}[!htb]
\begin{center}
\includegraphics[width=2.6in]{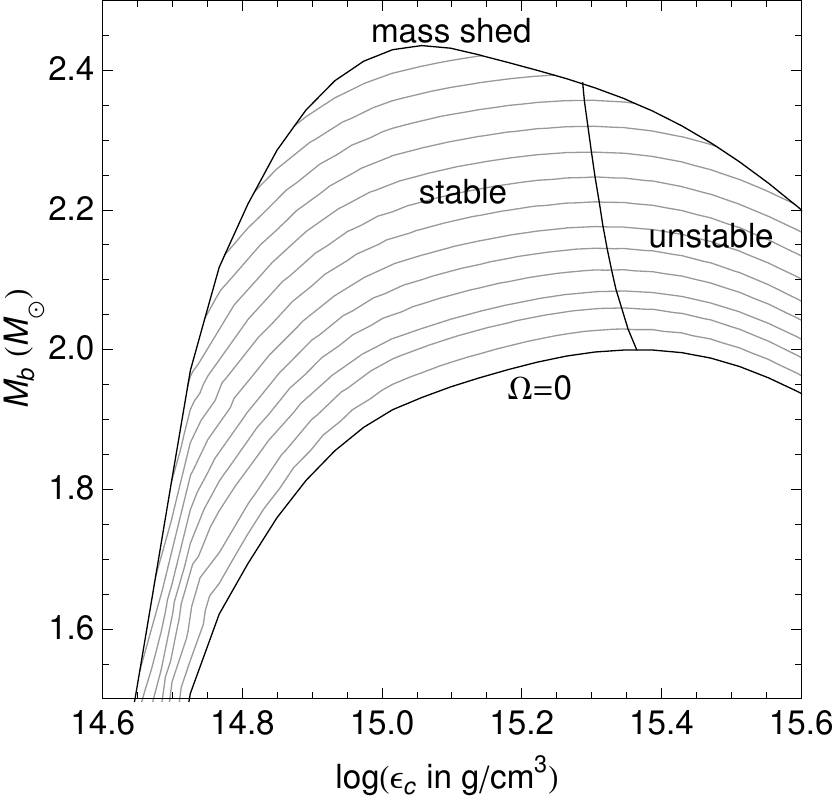}
\end{center}
\caption{Family of rotating models for a given EOS, from the spherical limit to
the mass-shed limit. The surface is projected into $M_b$-$\epsilon_c$ and
covered with lines of constant $J$.  The maximum $M_b$ on each line gives the
critically stable model; rotation at the mass-shed limit increases as
$\epsilon_c$ increases past the maximum mass model.  }
\label{fig:maxrotstab}
\end{figure}

For most $npe\mu$-only EOSs, the maximally rotating stable model is close to
the point on the mass-shed limit with maximal mass-energy, and this model has
been used for an estimate of maximal rotation in surveys of large numbers of
EOSs.  However, this is not always the case. An example is in EOS L of
\cite{CookShapiroTeukolsky}, or the parameterized EOS of Fig.
\ref{fig:maxrotstab}.  

Consider the two-parameter family of rotating neutron stars as a surface
$\Sigma$ in $M_b$-$J$-$\epsilon_c$ space. The central energy density
$\epsilon_c$ and axis ratio $\mathfrak{r}$ are suitable parameters for this
surface.  At points where $M_b$ is maximum along constant $J$ sequences, the
vector tangent to the sequence points in the $\epsilon_c$ direction.  Models
with limiting stability are found where the tangent plane to the surface of
equilibrium models contains a vector in the $\epsilon_c$ direction,
$\widehat{\bm \epsilon_c}$.  

Given the parameterization of the surface in terms of $\epsilon_c$ and
$\mathfrak{r}$, the normal vector to the surface ${\bm \Sigma} = \left\{ M_b,
J, \epsilon_c\right\}$ is along
\begin{equation}
{\bm n} = \frac{\partial\mathbf{\Sigma}}{\partial \epsilon_c} \times
\frac{\partial\mathbf{\Sigma}}{\partial \mathfrak{r}} 
\end{equation}
with component along ${\widehat{\bm \epsilon_c}}$.
\begin{equation}
\label{eq:rotstab}
n_{\epsilon_c} = 
\frac{\partial M_b}{\partial \epsilon_c} 
\frac{\partial J}{\partial \mathfrak{r}} 
- \frac{\partial J}{\partial \epsilon_c}
\frac{\partial M_b}{\partial \mathfrak{r}}
\end{equation}
which is zero at the critical line between stable and unstable equilibriums on
the surface $\bm{\Sigma}$.  A covariant statement of 
this condition for marginal stability is $dM_b\wedge dJ =0$.

The maximally rotating model for a given EOS may be determined, without finding
sequences of constant $J$ and $M_b$, by considering a sequence of central
energy densities $\epsilon_c$. First, increase the axis ratio $\mathfrak{r}$
until the Kepler limit is found, as in the example program \texttt{main.c} of
\texttt{rnsv2.0}. Second, vary $\epsilon_c$ and $\mathfrak{r}$ around this
point to estimate the partial derivatives of Eq.~\ref{eq:rotstab}.  The sign of
$n_{\epsilon_c}$ will change as the Kepler limit sequence crosses the stability
limit.

\section{Analytic fits to tabulated EOSs}\label{ap:paramacc}

As another measure of the ability of the parameterized EOS to fit candidate
EOSs from the literature, we examine how well the parameterized EOS reproduces
neutron star properties predicted by the candidate EOSs.  We use an analytic
form of the (SLy) low-density EOS that closely matches its tabulated values.
With rms residual less than $0.03$, $p(\rho)$ for SLy is approximated between
$\rho=10^3$~g/cm$^3$ and $\rho=10^{14}$~g/cm$^3$ by four polytropic pieces.
The four regions correspond roughly to a nonrelativistic electron gas, a
relativistic electron gas, neutron drip, and the density range from neutron
drip to nuclear density. Using the notation of Sect.~\ref{sec:ppdef}, the
analytic form of the SLy EOS is set by the values of $K_i, \Gamma_i$ and
$\rho_i$ listed in Table~\ref{tab:LowEOS}. The parameters for the three piece
polytropic high-density EOS, the corresponding residuals, as well as
the observable properties of these EOSs and the error in using the best fit
parameterized EOS instead of the tabulated EOS are listed in
Table~\ref{tab:constraints}.  The parameterized EOS
systematically overestimates the maximum speed of sound.

\begin{table}[!htb]
\caption{An analytic representation of $p(\rho)$ for the SLy EOS below nuclear
density uses polytropes specified by the constants listed here. $\Gamma_i$ is
dimensionless, $\rho_i$ is in $\mathrm{g/cm^{3}}$, and $K_i$ is in cgs units
for which the corresponding value of $p$ is in units of $\mathrm {dyne/cm^2}$.
The last dividing density is the density where the low density EOS matches the
high density EOS and depends on the parameters $p_1$ and $\Gamma_1$ of the high
density EOS.}
\label{tab:LowEOS}
\begin{center}
\begin{tabular}{ccc}
\hline\hline
$K_i$ & $\Gamma_i$ & $\rho_i$ \\
\hline
6.80110e-09 & 1.58425 & 2.44034e+07\\
1.06186e-06 & 1.28733 & 3.78358e+11\\
5.32697e+01 & 0.62223 & 2.62780e+12\\
3.99874e-08 & 1.35692 & --\\
\hline\hline
\end{tabular}
\end{center}
\end{table}

\begin{table*}[!htb]
\begin{center}
\caption{
Comparison of candidate EOSs and their best fits.  The parameters that
provide the best fit to the candidate EOSs as well as the residual are
given. $p_1$ is in units of dyne/cm$^2$.
Values for observables calculated using the tabulated EOSs are also
given. $v_{\rm s,max}$ is the maximum adiabatic speed of sound
below the central density of the maximum mass neutron star.  $M_{\rm max}$
is the maximum nonrotating mass configuration in units of $M_\odot$.
$z_{\rm max}$ is the corresponding maximum gravitational redshift.  $f_{\rm
max}$ is the maximum rotation frequency in Hz, as calculated using the
rotating neutron-star code \texttt{rns}.  $I_{1.338}$ is the moment of
inertia for a $1.338~M_\odot$ star in units of $10^{45}$~g\,cm$^2$.
$R_{1.4}$ is the radius of a $1.4~M_\odot$ star in units of km.  The
difference in calculated values for each observable when using the
tabulated EOS $(O_{\rm tab})$ versus the best fit parameterized EOS
$(O_{\rm fit})$ is calculated with
$(O_{\rm fit}/O_{\rm tab}-1)100$.  The last entry gives the mean and standard
deviation of the errors for each observation.} 
\label{tab:constraints}
\begin{tabular}{l|ccccl|rr|rr|rr|rr|rr|rr}
\hline\hline
EOS & $\log(p_1)$ & $\Gamma_1$ & $\Gamma_2$ & $\Gamma_3$ & residual &
$v_{s, \rm max}$ & \% & $M_{\rm max}$ & \% & $z_{\rm max}$ & \% & $f_{\rm max}$ & \% & $I_{1.338}$ & \% & $R_{1.4}$ & \%\\
\hline
PAL6&34.380&2.227&2.189&2.159&0.0011& 0.693&1.37& 1.477&-0.47& 0.374&-0.51& 1660&-0.97& 1.051&-2.03& 10.547&-0.54\\
SLy&34.384&3.005&2.988&2.851&0.0020& 0.989&1.41& 2.049&0.02& 0.592&0.81& 1810&0.10& 1.288&-0.08& 11.736&-0.21\\
AP1&33.943&2.442&3.256&2.908&0.019& 0.924&9.94& 1.683&-1.60& 0.581&2.79& 2240&1.05& 0.908&-2.57& 9.361&-1.85\\
AP2&34.126&2.643&3.014&2.945&0.0089& 1.032&0.42& 1.808&-1.50& 0.605&0.33& 2110&-0.02& 1.024&-2.34& 10.179&-1.57\\
AP3&34.392&3.166&3.573&3.281&0.0091& 1.134&2.72& 2.390&-1.00& 0.704&0.57& 1810&-0.14& 1.375&-1.59& 12.094&-0.96\\
AP4&34.269&2.830&3.445&3.348&0.0068& 1.160&1.45& 2.213&-1.08& 0.696&0.22& 1940&0.05& 1.243&-1.36& 11.428&-0.90\\ 
FPS&34.283&2.985&2.863&2.600&0.0050& 0.883&2.29& 1.799&-0.03& 0.530&0.67& 1880&0.11& 1.137&0.03& 10.850&0.12\\
WFF1&34.031&2.519&3.791&3.660&0.018& 1.185&7.86& 2.133&-0.29& 0.739&2.21& 2040&0.30& 1.085&0.10& 10.414&0.02\\ 
WFF2&34.233&2.888&3.475&3.517&0.017& 1.139&7.93& 2.198&-0.14& 0.717&0.71& 1990&0.03& 1.204&-0.59& 11.159&-0.28\\
WFF3&34.283&3.329&2.952&2.589&0.017& 0.835&8.11& 1.844&-0.48& 0.530&2.26& 1860&0.59& 1.160&-0.25& 10.926&-0.12\\
BBB2&34.331&3.418&2.835&2.832&0.0055& 0.914&7.75& 1.918&0.10& 0.574&0.97& 1900&0.47& 1.188&0.17& 11.139&-0.29\\
BPAL12&34.358&2.209&2.201&2.176&0.0010& 0.708&1.03& 1.452&-0.18& 0.382&-0.29& 1700&-1.03& 0.974&0.20& 10.024&0.67\\
ENG&34.437&3.514&3.130&3.168&0.015& 1.000&10.71& 2.240&-0.05& 0.654&0.39& 1820&-0.44& 1.372&-0.97& 12.059&-0.69\\
MPA1&34.495&3.446&3.572&2.887&0.0081& 0.994&4.91& 2.461&-0.16& 0.670&-0.05& 1700&-0.18& 1.455&-0.41& 12.473&-0.26\\
MS1&34.858&3.224&3.033&1.325&0.019& 0.888&12.44& 2.767&-0.54& 0.606&-0.52& 1400&1.67& 1.944&-0.09& 14.918&0.06\\
MS2&34.605&2.447&2.184&1.855&0.0030& 0.582&3.96& 1.806&-0.42& 0.343&2.57& 1250&2.25& 1.658&0.46& 14.464&-2.69\\
MS1b&34.855&3.456&3.011&1.425&0.015& 0.889&11.38& 2.776&-1.03& 0.614&-0.56& 1420&1.38& 1.888&-0.64& 14.583&-0.32\\
PS&34.671&2.216&1.640&2.365&0.028& 0.691&7.36& 1.755&-1.53& 0.355&-1.45& 1300&-2.39& 2.067&-3.06& 15.472&3.72\\
GS1\footnotemark[1]&34.504&2.350&1.267&2.421&0.018& 0.695&0.49& 1.382&-1.00& 0.395&-0.64& 1660&9.05& 0.766&-3.13& \footnotemark[2]&\\
GS2\footnotemark[1]&34.642&2.519&1.571&2.314&0.026& 0.592&16.10& 1.653&-0.30& 0.339&7.71& 1340&3.77& 1.795&-3.33& 14.299&0.07\\
BGN1H1&34.623&3.258&1.472&2.464&0.029& 0.878&-7.42& 1.628&0.39& 0.437&-3.55& 1670&-2.08& 1.504&0.56& 12.901&-1.96\\
GNH3&34.648&2.664&2.194&2.304&0.0045& 0.750&2.04& 1.962&0.13& 0.427&0.37& 1410&-0.04& 1.713&0.38& 14.203&-0.28\\
H1&34.564&2.595&1.845&1.897&0.0019& 0.561&2.81& 1.555&-0.92& 0.311&-1.47& 1320&-1.46& 1.488&-1.45& 12.861&-0.03\\
H2&34.617&2.775&1.855&1.858&0.0028& 0.565&1.38& 1.666&-0.77& 0.322&-0.55& 1280&-1.29& 1.623&-0.82& 13.479&0.29\\
H3&34.646&2.787&1.951&1.901&0.0070& 0.564&7.05& 1.788&-0.79& 0.343&1.07& 1290&-0.88& 1.702&-1.18& 13.840&0.31\\
H4&34.669&2.909&2.246&2.144&0.0028& 0.685&4.52& 2.032&-0.85& 0.428&-1.01& 1400&-1.28& 1.729&-1.18& 13.774&1.34\\
H5&34.609&2.793&1.974&1.915&0.0050& 0.596&1.65& 1.727&-1.00& 0.347&-0.82& 1340&-1.55& 1.615&-1.31& 13.348&0.68\\
H6\footnotemark[1]&34.593&2.637&2.121&2.064&0.0087& 0.598&11.71& 1.778&0.07& 0.346&8.65& 1310&5.33& 1.623&-2.19& 13.463&0.37\\
H7&34.559&2.621&2.048&2.006&0.0046& 0.630&1.82& 1.683&-1.12& 0.357&-0.57& 1410&-1.52& 1.527&-2.33& 12.992&0.23\\
PCL2&34.507&2.554&1.880&1.977&0.0069& 0.600&1.74& 1.482&-0.79& 0.326&-2.25& 1440&-1.87& 1.291&-3.27& 11.761&-1.15\\
ALF1&34.055&2.013&3.389&2.033&0.040& 0.565&18.59& 1.495&-0.53& 0.386&3.52& 1730&2.44& 0.987&-0.40& 9.896&-0.22\\
ALF2&34.616&4.070&2.411&1.890&0.043& 0.642&1.50& 2.086&-5.26& 0.436&-0.62& 1440&1.01& 1.638&-6.94& 13.188&-3.66\\
ALF3&34.283&2.883&2.653&1.952&0.017& 0.565&11.29& 1.473&-0.06& 0.358&2.46& 1620&1.79& 1.041&0.76& 10.314&-0.25\\
ALF4&34.314&3.009&3.438&1.803&0.023& 0.685&14.78& 1.943&-0.93& 0.454&0.59& 1590&0.52& 1.297&-2.38& 11.667&-1.20\\
\hline
\multicolumn{6}{l|}{Mean error} && 5.68 && -0.71 && 0.74 && 0.43 && -1.26 && -0.37\\
\multicolumn{6}{l|}{Standard deviation of error} && 5.52 && 0.96 && 2.42 && 2.25 && 1.57 && 1.29\\
\hline\hline
\end{tabular}
\footnotetext[1]{The tables for GS1, GS2, and H6 do not go up to the central
density of the maximum mass star.  For most observables, the EOS can be safely
extrapolated to higher density with minimal error.  However, the maximum speed
of sound is highly sensitive to how this extrapolation is done.  Thus, we only
use the maximum speed of sound up to the last tabulated point when comparing the
values for the table and fit.}
\footnotetext[2]{GS1 has a maximum mass less than $1.4~M_\odot$.}
\end{center}

\end{table*}

\end{document}